\documentclass[a4]{article}

\usepackage{algorithm}
\usepackage[noend]{algpseudocode} 

\usepackage{moreverb,url}
\usepackage[utf8]{inputenc} 
\usepackage[T1]{fontenc}    
\usepackage{booktabs}       
\usepackage{amsfonts}       
\usepackage{nicefrac}       
\usepackage{comment}
\usepackage{float}
\usepackage{graphicx}
\usepackage{caption}
\usepackage{subcaption}
\usepackage[export]{adjustbox}
\usepackage{supertabular}
\graphicspath{ {IMAGES/} }
\usepackage{amsmath,amssymb}
\usepackage{changepage}
\usepackage{textcomp,marvosym}
\usepackage{cite}
\usepackage{nameref,hyperref}
\usepackage[right]{lineno}
\usepackage[table]{xcolor}
\usepackage{array}

\newcolumntype{+}{!{\vrule width 2pt}}

\newlength\savedwidth

\usepackage[aboveskip=1pt,labelfont=bf,labelsep=period,
singlelinecheck=off]{caption}

\makeatletter
\renewcommand{\@biblabel}[1]{\quad#1.}
\makeatother

\usepackage{lastpage,fancyhdr,graphicx}
\usepackage{epstopdf}
\fancyheadoffset[L]{2.25in}
\newcommand{\luis}[1]{\textcolor{blue}{[luis: #1]}}
\newcommand{\wolf}[1]{\textcolor{red}{[wolf: #1]}}

\begin{document}
\vspace*{0.2in}

\definecolor{myEcoliColor}{HTML}{ffe187}
\definecolor{myBsubColor}{HTML}{c7b9ff}


\begin{flushleft}
{\Large
\textbf{Complexity reduction by symmetry: uncovering the minimal regulatory network for logical computation in bacteria} 
}
\newline
\\

Luis A. Álvarez-García\textsuperscript{1\ddag},
Wolfram Liebermeister\textsuperscript{2},
Ian Leifer\textsuperscript{1},
Hern\'an A. Makse\textsuperscript{1*}
\\
\bigskip    
\textbf{1} Levich Institute and Physics Department, City College of New York, New York, NY 10031, USA
\\
\textbf{2} Universit\'e Paris-Saclay, INRAE, MaIAGE, 78350 Jouy-en-Josas, France
\\
\bigskip

%
%





\ddag luisalvarez.10.96@gmail.com
* hmakse@ccny.cuny.edu

\end{flushleft}

\section*{Abstract}
Symmetry principles play an important role in geometry, and physics, allowing for the reduction of complicated systems to simpler, more comprehensible models that preserve the system's features of interest. Biological systems are often highly complex and may consist of a large number of interacting parts. Using symmetry fibrations, the relevant symmetries for biological "message-passing" networks, we introduce a scheme, called
Complexity Reduction by Symmetry or ComSym, to reduce the gene regulatory networks of \emph{Escherichia coli} and \emph{Bacillus subtilis} bacteria to core networks in a way that preserves the dynamics and uncovers the computational capabilities of the network. Gene nodes in the original network that share isomorphic input trees are collapsed by the fibration into equivalence classes called fibers, whereby nodes that receive signals with the same "history" belong to one fiber and synchronize. Then we reduce the networks to its minimal computational core via k-core decomposition. This computational core consists of a few strongly connected components or "signal vortices", in which signals can cycle through. While between them, these "signal vortices" transmit signals in a feedforward manner. These connected components perform signal processing and decision making in the bacterial cell by employing a series of genetic toggle-switch circuits that store memory, plus oscillator circuits.  These circuits act as the central computation device of the network, whose output signals then spread to the rest of the network. Our reduction method opens the door to narrow the vast complexity of biological systems to their minimal parts in a systematic way by using fundamental theoretical principles of symmetry.


\section*{Author summary} 
Biological systems are constituted by complex interactions between a large number of different components, and being able to reduce their complexity in order to understand their behavior is of paramount importance. Here we use symmetry principles, in a manner akin to physics, to reduce the Gene Regulatory Networks (GRN) of \emph{Escherichia coli} and \emph{Bacillus subtilis} bacteria to reveal the computational core structure of these networks responsible for driving their dynamics.  This computational core comprises gene logic circuits, such as toggle-switches and oscillatory circuits, which ultimately are in charge of the decision making in the bacterial cell. This Complexity Reduction by Symmetries (ComSym) method opens the way to understanding biological complexity based on firm theoretical principles.


\textbf{Keywords:} graph fibration, gene regulatory network, gene logic circuit, toggle-switch, Escherichia coli, Bacillus subtilis, k-core decomposition, network motif, fibers, cluster synchronization, fiber building blocks, simple directed cycles

\textbf{Abbreviations:} GRN: Gene regulatory network; SSC: Strongly Connected Component; ODE: Ordinary differential equation; TU: Transcriptional Unit; TF: transcription factor; NAR: negative auto-regulation; PAR: positive auto-regulation; MR: mutual repression; NFBL: negative feedback loop

\section{Introduction}

\paragraph{Complexity reduction through symmetries}

One of the main challenges of systems biology is that biological systems are inherently complex, as often reflected in the sheer mass of quantitative parameters and details needed to describe such systems accurately and precisely~\cite{book-wolf}. The human brain, an evident example, consists of $\sim$ 80 billion neurons with 100 trillion connections between them, each one
with a large set of parameters defining the strength of their
interactions. In the mouse brain, with three orders of magnitude fewer
neurons, advanced techniques need to be employed to understand the collective macroscopic behavior~\cite{meshulam2019coarse}. Even in the neural system of {\it C. elegans} worms, composed of merely 302 neurons, it is not known how this tiny
connectome leads to function, and low-dimensional models are needed
and regularly used~\cite{varshney2011structural,morone2019symmetry}.

High-dimensional parameter spaces are ubiquitous in biological systems. Finding low-dimensional effective models to describe the dynamics of these systems is crucial to understand how function and collective behavior emerge from the complex dynamics of the system's constitutive elements. This is where concepts and methods from physics have proven to be of great help~\cite{bialek2012statistical, berman2014mapping, stephens2008dimensionality, stephens2010modes}. In physical systems, one often encounters the challenge of handling high-dimensional experimental data. Fortunately, solid theoretical methods have been developed to address this, one of the most powerful being the use of symmetries. 

Fibration symmetries~\cite{boldi2002fibrations, morone2020fibration} 
are symmetries, or redundancies, of the pathways through which signals, or messages, are transmitted, as will be explained further in Section~\ref{sec: Fibrations}. These symmetries are identified by finding sets of nodes with identical input trees, known as fibers. Grouping nodes into fibers is particularly useful because it allows for the reduction of network complexity by collapsing symmetric nodes, without disrupting the network's "information flow". 
This reduction has been applied for understanding the structure of gene regulatory networks (GRNs) in bacteria, enabling us to simplify these networks and present a more transparent understanding while preserving their dynamics~\cite{morone2020fibration, 2020predict}. 

Furthermore, a partition of the network's nodes into its fibers allows for the detailed breakdown of these networks into their constitutive \emph{building block} components, their canonical forms in Fig.~\ref{fig:circuitsOverview} as presented in Ref~\cite{morone2020fibration} with examples taken from \emph{E. coli}'s GRN. 


\begin{figure*}[!t] 
    \centering
    \includegraphics[width=\textwidth]{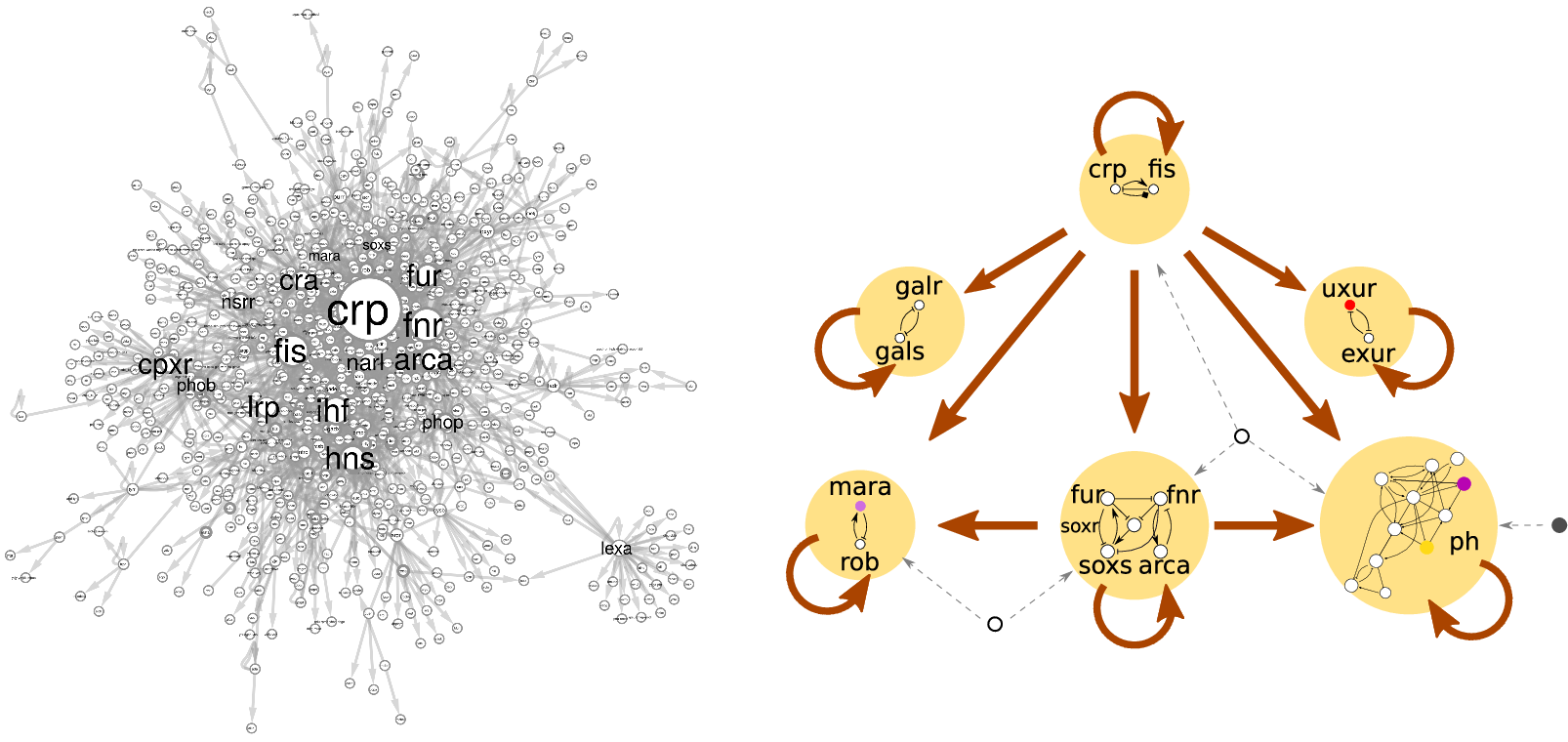}
\caption{
{\bf A minimal GRN obtained through the ComSym reduction method}. 
Reduction of Gene Regulatory Network (GRN) of \emph{E.~coli} bacteria. Left: 879 nodes operon-GRN of \emph{E.~coli}. Here we show only the weakly connected
  component, i.e.small disconnected pieces of the network are not shown, since
  they do not play a significant role for the network dynamics. Node sizes and font size are proportional to out-degree of the
  node.  Right: a representative of the minimal GRN obtained
  after the application of our ComSym reduction method. The simple depiction of the core network illustrates the signal flow between its different components
  (bigger "nodes"), the strongly connected components of the
  network (SCC: a component of a graph that within it, each node can be reached from every other node). The smaller gene nodes inside these SCCs form the computational core of the
  network. Colored nodes represent collapsed-fiber nodes. Bigger arrows represent the edges between the components. The three nodes
  outside the components are representative of controller nodes, 
  which send signals to different SCCs. Interestingly,
  two parallel feedforward structures exist between the components: the
  central \emph{crp-fis} SCC regulates the \emph{soxS} SCC. In one feedforward structure, they regulate jointly the pH SCC and in the other one the \emph{mara-rob} SCC. 
  }
\label{fig:main}
\end{figure*}

\paragraph{Gene regulatory networks (GRNs)}
Here we focus on the regulation of bacterial genes as an example of signal processing in a complex biological system. The model bacterium \emph{E.~coli} for example, possesses a genome of more than 4,000 genes (compiled by RegulonDB's aggregate of results to date~\cite{tierrafria2022regulondb}), 1843 of which are known to regulate other genes through transcription factor (TF) proteins.

The \emph{Gene Regulatory Network} (GRN) is made up of all the transcriptional regulations between genes.
When a gene regulates another, via a TF protein that binds to the regulated gene's binding region, this constitutes a directed edge in the GRN between the two (genes) nodes. 
However, every individual transcriptional interaction between two genes requires a multitude of microscopic parameters for a precise mathematical description of the gene expression dynamics~\cite{book-wolf}. These parameters include everything from transcription and translation rates to binding and unbinding of the TFs as well as ribosomes, for example.
In order to accurately describe the global dynamics of the GRNs through ordinary differential equations (ODEs), all of these microscopic parameters would need to be known.

Given that most of these parameters are unknown, structural analyses of the GRNs formed by these "lumped" edges have typically been forced to overlook these heterogeneities, treating all edges representing the same set of parameters and modulo up-regulation and down-regulating effects. In effect, this simplifies the system to two main regulatory types, distinguishing only between two main types, namely, repressors and activators. We continue with this approach; given that it has led to considerable insights in the past, we refer to Section~\ref{sec:SI:GeneRegulationDynamics} of the Supplementary Information for more details on these challenges. 

In particular, the discovery of network motifs, small and local motifs that have been identified by statistical overrepresentation in the network,
compared to randomized networks that preserve the same observed degree distribution~\cite{shen2002network, milo2002network}. 
Although the individual dynamics of some network motifs is more or less
understood~\cite{mangan2003structure, shen2002network, milo2002network}, given that they are in essence local structures, they do little to unravel the global
topology of the network or the global dynamics. 

On a large scale, it has been proposed that the {\it E.~coli}'s GRN has a
feedforward structure~\cite{thieffry1998specific, ma2004hierarchical, martinez2008functional, dobrin2004aggregation,shen2002network}, where signals flow unidirectionally from a core of sensors and master regulators
through a series of parallel layers down to an outer periphery in
a feedforward manner \cite{dobrin2004aggregation, shen2002network, milo2002network}. The modular structure of the network has also been noted~\cite{dobrin2004aggregation}.
However, there also exist a significant number of feedback loops, which complicate this picture. Thus, many questions remain unanswered. What is the relevant structure at the "center" of these systems. What is the core structure responsible for decision making? Which genes belong to this computational core? How does this structure control the rest
of the network? Is there a minimal computation core that explains the
structure and function of the GRN in a simplified manner?

\paragraph{Gene regulatory networks as a computing device}

We can interpret a transcriptional regulation from one gene to another as a form of "transcriptional signal" that one gene sends to another. Representing a regulatory "message" with the TF as a "messenger". The pathways of the GRN can then be seen as "signaling pathways", on aggregate establishing the "signaling flow", or "information" (loosely stated) flow in the network.
A main feature of GRNs, which we emphasize here, is that signals do not only propagate in "forward direction" between different layers of the network, but can also cycle in feedback loops. This is significant because network structures that only allow forward transmission, or sequential logic, map input signals to output signals of similar shapes, possibly blurred, inverted, or time-delayed~\cite{book-wolf}. They may also aggregate several inputs and generate several outputs, but it is not possible for them to have a memory of previous states. Circuits in which signals can cycle show more complex behavior: they can stabilize an output variable, generate oscillations, and may internally store information, like a toggle switch in synthetic biology~\cite{gardner2000construction} analogous to a flip-flop in electronics.

In electronics, feedforward circuits are called "combinatoric logic
circuits" and are memoryless digital circuits whose output at a given
time depends only on the combination of its inputs. These circuits are
made of standard logic gates such as NOR and NAND. On the other hand,
circuits with feedback are called "sequential logic circuits." Their
output depends on both their present input and also their previous
output. This feedback loop provides them with memory, since the
circuit is able to "remember" its state even when the external input is
removed. 
Combinatorial and sequential circuits are the "decision-making" machinery behind the logic function of electronic circuits. 

{
Even though a gene's expression level is in fact a continuous quantity, a Boolean logic approximation can help picture it's expression levels. Indeed, a GRN modeled under a Boolean approximation can simulate any finite state machine~\cite{oishi2014framework} (a simple sequential computing device with memory of its state).}
Furthermore, given the important role of GRNs in the regulation of the reaction of bacteria to both its internal and external states, it is reasonable to assume that they would play a central role in their \emph{decision-making} dynamics. 
So, if the cell is to perform as a form of biological computational device (in the sense that it needs to "output" a reaction in response to the external "inputs" it encounters and its internal "states"), then both forms of sequential and combinatoric logic circuits should play an important role in the GRN. Indeed, we find these two modes of signal transmission appear both on a small and on a large scale in the GRN.


\begin{figure}[t!] 
   \begin{center}
    \includegraphics[width=1\textwidth]{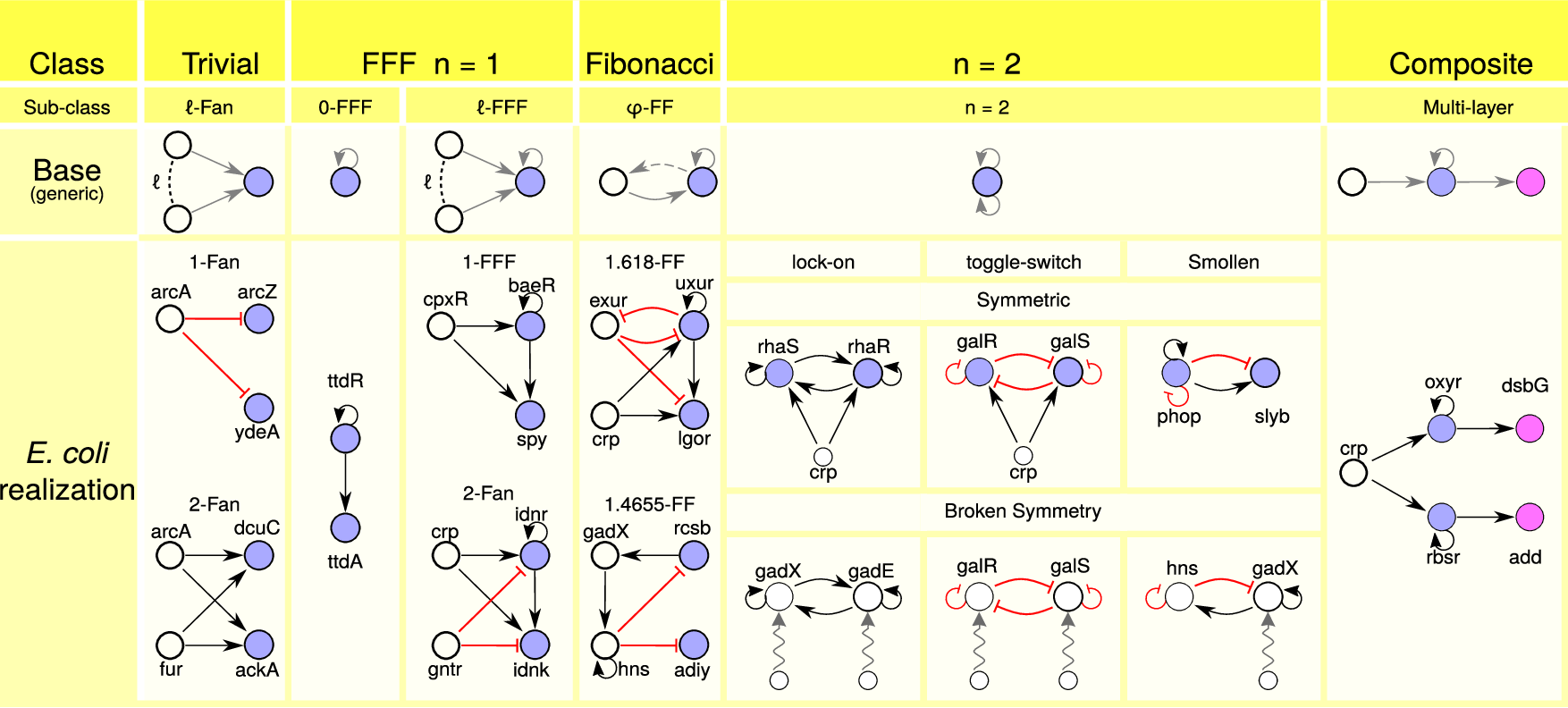}
    \end{center}
    \caption{
    {\bf Canonical fiber building blocks.} These correspond to the canincal fiber building blocks observed in the GRNs of \emph{E.~coli} and \emph{B.~subtilis}, with examples taken from \emph{E. coli}. 
    The networks can be seen as assemblies of 5 basic classes of fibration building blocks:
      (i) Trivial fibers. A number $\ell$ of external regulators
      identically regulate the genes in a fiber, which then show synchronous dynamics.  Operons with only one promoter belong to this class, where colored nodes
      represent genes belonging to the operon (perhaps with more
      colored nodes in the fibers, depending on the number of genes in
      the operon). (ii) The feedforward fiber 
      and its sub-classes of $\ell$-FFF with $\ell$ external regulators. The
      FF fiber is defined by a feedforward motif with a self-loop in
      the synchronous set of genes, and the number of $\ell$ external
      regulators. (iii) The Fibonacci fiber, $\varphi$-FF. A more
      complex building block, defined by a fractal dimension branching
      ratio that occurs given the presence of a self-loop and a
      feedback regulation from the
      fiber back to the regulator(s). The Fibonacci fibers observed  in \emph{E.~coli} have a
      branching ratio between 1 and 2, placing this building block in
      between the FFF fibers and the n=2 fibers. (iv) The n=2 fibers,
      defined by two self-loops in the synchronized genes. When this
      symmetry is broken it forms the memory and oscillatory logic circuits
      embedded in the SCCs. And finally (v) composite fibers of the
      previous ones. By adding different types of the previous 4
      building blocks, in a sequential manner, a composite fiber is
      obtained. An interesting consequence of this is the
      synchronization of genes that may be far apart from each other
      and don't share any regulation. 
      }
    \label{fig:circuitsOverview}
\end{figure}

\paragraph{Complexity reduction through fibrations: constructing a minimal TRN}
In this work, we describe how to reduce the complexity of GRNs to their "minimal computational core" by applying the novel tool of fibration symmetries, in addition to standard graph-theoretical tools. Based on this notion of symmetry, we present a novel method to reduce any directed \emph{"message-passing"} network (any network where edges represent signals between nodes) to what we call its computational core. After the GRN is reduced to its minimal structure, we analyze this minimal structure. The
overall procedure consists of five steps.

Steps (I) and (II) are concerned with the reduction of the network, removing all its elements that do not contribute to its computational capabilities. Reducing the GRN to the core network at the heart of the decision-making processes: the minimal GRN. Step (I) eliminates all redundant information pathways 
through the use of graph fibrations. While step (II) removes the nodes that only receive signals or just pass-it-through without contributing to the decision-making process via the $k$-core decomposition, these nodes are responsible for communicating the output from the minimal GRN to the peripheries. Steps (III)-(V) analyze this minimal GRN. In step (III) we focus on the large-scale structure of the minimal GRN: how the components of this core network are connected with each other. The last two steps "zoom-in" to look at the small-scale, or local, structures within the minimal GRN's components, by looking at the logic circuits (step IV) and how these are connected with each other as well as informing about the connectivity structure within the different components (step V).
  
Steps (I) and (IV) decompose the genetic network into its building blocks by using fibration symmetries (Fig.~\ref{fig:circuitsOverview}) and broken symmetries, respectively. The fiber building blocks correspond to the 3 basic canonical types shown in Fig.~\ref{fig:circuitsOverview}: I) simple $n$-ary trees $\vert n, \ell\rangle$, II) Fibonacci building blocks and III) Composite Building Blocks. 

\paragraph{The structure of the minimal GRN: Large-scale components and small-scale, local circuits}
We apply our method to model bacteria \emph{E.~coli} and \emph{B.~subtilis}. In the case of \emph{E.~coli}, reducing it from the entire GRN to the left in Fig.~\ref{fig:main} to the much simpler network to the right (representing a sketch of its computational core). 
The resulting minimal GRNs correspond to just a tiny fraction of the total genes and are composed of vortices on a large scale and feedforward and feedback circuits on a small scale. Recognizing this structure helps us to understand what the network can do and what the functions are of different parts of the network. Hence, we not only reduce the GRN to its core computational structure (by omitting symmetric and peripheral nodes), but also identify the "signal vortices" and the local gene circuits that process signals and store information and are, therefore, capable of computations, comparable to a silicon-based computer.

On a large-scale level, the core network of any directed network consists of Strongly Connected Components (SCCs) connected between them in a "forward signaling" manner, along with the nodes that regulate them. See, for example, the right panel of Fig.~\ref{fig:main}, which represents the core network at the heart of \emph{E.~coli}'s GRN: the minimal GRN. The SCCs correspond to "signal vortices" in which signals can cycle and then gets sent to the fibers outwardly of this minimal GRN to the periphery. Embedded within these SCCs are the small-scale structures, the logic circuits of the GRN: symmetry-broken memory-storing
toggle-switches and oscillators; the two primary components of any
computer~\cite{tanenbaum2016structured}. 
The system of SCCs represents the smallest computational subunits that cannot be further reduced by fibration symmetries or by smaller strongly connected components. Thus, this structure represents the "minimal GRN" structure of the cell. 
Thus, the GRN can be seen as a computational machine, where the memory is stored and controlled by broken symmetry circuits within the SCCs.


\section{Methods}

\subsection{Symmetries and fibrations}



Symmetries play a crucial role in physics by simplifying complex models. Noether's theorem established a profound connection between the symmetries of a Lagrangian and its conserved physical quantities in classical mechanics, such as energy and momentum. By identifying and exploiting these symmetries, physicists can reduce the number of variables needed to describe a system while preserving its essential characteristics. This approach has become ubiquitous in physics when dealing with complex problems, from classical mechanics to quantum mechanics and particle physics, where symmetries underpin the Standard Model under the gauge symmetry group U(1)$\times$SU(2)$\times$SU(3), for example.


Motivated by the success of applying symmetries and geometry across various fields, we ask whether a similar approach can be used to tame biological complexity and, if so, in what specific way. For example, what are the symmetries of a system such as \emph{E.~coli's} GRN, left panel in Fig.~\ref{fig:main}? As we will explore, symmetries can help us understand these systems, but with an important distinction: the symmetry groups used in physics cannot be directly applied to biological systems. Instead, a new type of symmetry, based on \emph{graph fibrations}~\cite{boldi2002fibrations, deville2013modular}, turns out to be more suitable.

Since we are dealing with networks, the initial assumption would be to look at automorphisms, 
transformations of a graph that leave it unchanged.
However, automorphisms are too restrictive for understanding biological signaling networks \cite{morone2020fibration, morone2019symmetry, leifer2020circuits, 2020predict}. The symmetries that have been so fundamental in physics do not translate well to the complexity of biochemical systems, necessitating a different approach.

Why, then, do automorphisms fail in biology? One could argue that a fully symmetric graph, while significant in physics, 
maybe less useful to represent a biological system on account of its sensitivity to small alterations which are commonplace in biology. 
A biological signaling network captures the flow of information dictated by the dynamic interactions between biological units—whether neurons in the brain, genes, or enzymes in a genetic or metabolic network. The information that reaches each node is determined by its inputs alone, not its outputs. Consequently, the full structure (including both input and output) is not crucial for understanding the function of a node in these networks.

In our particular case of gene networks, the flow of signals is well defined. Because the dynamics of each individual component is only determined by its inputs, this begs the question: What if instead we focus only on the inputs of a node? The "history" of the signals, transmitted through the pathways of the network, or input trees, and their symmetries.

As we will explain, nodes with identical input trees will, in fact, share the same dynamical state, a form of symmetry. In other words, nodes that share identical input "history" are symmetric in a certain sense 
and become synchronized. This is significant because cluster synchronization is ubiquitous across biological systems, from gene expression~\cite{van2018gene, stuart2003gene} to brain activity~\cite{bialek2012statistical,meshulam2019coarse}. The overall function of cells must be driven in part by coherent communication between their units, and this synchronization is captured by the fibration symmetries of the underlying biological graph, as we will show. Thus, the existence of symmetries in the information pathways of the network
helps explain and predict the function of the network, linking its structure to its behavior.

\subsection{Gene Regulatory Networks (GRNs)}

Gene expression in cells is regulated in response to the cells' environment and internal state. In bacteria, gene expression is governed by global mechanisms (e.g.~via sigma factors, varying activity of the global transcription and translation machinery), where also "regional effects", such as the chromosome correlated expression of nearby genes on the chromosome, play a role. However, an important part of gene regulation in  bacteria occurs through transcription factors (TFs), proteins that bind to DNA sites called promoters that regulate specific genes. 

Transcription factors can activate or repress the transcription of genes, thus increasing or decreasing the expression levels of their target genes. The transcription factors activities depend on their own expression levels and can be modulated by small-molecule binding. By binding  to different TFs, metabolites can modify their activities and thus modulate the expression levels of their target genes. The metabolites may come from the outside environment of the cell or may be products of the metabolism of the cell itself, providing a feedback mechanism in which the metabolic state of the cell informs the regulation of transcription \cite{book-wolf, alon2019introduction, martinez2008functional}.

The Gene Regulatory Network (GRN), here understood in a narrow sense, represents the regulatory effects between Transcription Factors (TFs) by regulation of transcription. In the network, if a TF (encoded by gene A) can bind to the promoter region of a gene B and regulate its expression, this is represented by a directed link from gene A to gene B. The type of edge corresponds to the type of regulation being performed by the transcription factor, being activation, inhibition, or dual (depending on how the TF binds to the promoter region).  Together, all such regulations form the GRN, which determines the expression of individual genes according to the cell's sensed environment and its own internal, for instance metabolic, states. 


We can see here how the gene regulatory system needs to "read" inputs from the environment so that it can then "respond" in an appropriate manner, response that needs to be "signaled" to the rest of bacteria's components in order to enact the appropriate reaction. This system helps us elucidate the needs that bacteria have for "decision making", which will be an important idea in this work. This process of sending regulatory messages as signals across genes defines the information flow in the gene network. The signal can be though of as an "information package" or "message passing" being sent from the source gene to the target gene. See SI Section~\ref{sec:SI:GeneRegulationDynamics} for a more thorough explanation.


\subsection{Graph fibration formalism}
\label{sec: Fibrations}


The network reduction method (Steps I and II) explored in this paper is based on graph
fibrations. Fibrations were introduced by
Grothendieck~\cite{grothendieck1959technique} in the context of
category theory and algebraic geometry. Although the original work
applies to fibrations between categories and it remains a bit obscure
for pedestrians, fortunately, this work has been adapted
to graph fibrations by Boldi and Vigna~\cite{boldi2002fibrations} in
computer science. Their inspiration was a distributed system of
computer processors that need to be synchronized in clusters in a
coherent manner for proper global updates, as there is no point to
have a processor waiting for its update while being out of sync with the rest. A
computer system seen as a graph of processors with fibration
symmetries then guarantees coherent optimal processing.  As stated in
the illuminating words in Vigna's blog on fibrations
\url{https://vigna.di.unimi.it/fibrations}:

{\it ``If a graph $G$ is used to represent the structure of a network that exchanges messages, and the processors of the network execute the same algorithm starting from the same initial state, the existence of a fibration $\varphi: G \to H$ implies that, whatever algorithm is used, there are executions in which the behavior of the nodes in $G$ is fibrewise constant (i.e., all processors in the same fiber are always in the same state).''}

The concept of fibrations can tell us about the signal processing dynamics and synchronization in networks based on network structure alone~\cite{deville2013modular}. 
As such, it is crucial to understand cluster synchronization~\cite{Sorrentino:16, Pecora:14} (gene coexpression in our case) and the signal processing tasks performed by these networks.

Under some ideal assumptions, this idea can be directly translated into GRNs to help us understand how it might function as a computing device.
This was enough inspiration for us to look for fibrations in these systems in the first place~\cite{morone2020fibration}.

Concretely, the condition that every processor executes the same
algorithm is translated to the GRN as the condition that every input
function has the same parameters (every edge represents the same
equation, module repressor/activator type). This condition is natural
for computer processors, but controversial for biology, as discussed
above and in more detail in the SI Section~\ref{sec:SI:GeneRegulationDynamics}. Still, as usually done by physicists (inspired by the metaphor
of the spherical cow, the legend of the Gordian knot, and Occam's
razor), we translate this simplification to biology and try to understand the consequences later. 
Alas, the cow is not spherical, but in the absence of the "perfect" approach (or any approach at all), it is better to start with a sphere and then introduce details and refinement, as needed. Otherwise, we might risk losing the forest for being too concerned with how the leaves of the tree look.

{
Another assumption is that different pathways do not experience significant communication delays which would cause asyncronicity. Additionally, we have a Boolean logic approximation in mind throughout. Its important to note that, unless the input functions (and microscopic parameters therein) drastically change from gene to gene, when these assumptions are not met, our approach is not broken but instead of a \emph{"fiberwise constant"} behaviour (i.e. gene clusters turning "on" or "off" in unison), we would expect to observe gene coexpression levels. In fact, we have found that fibration symmetries are actually able to predict gene coexpression (or correlated) patterns~\cite{2020predict}.
}


\paragraph{Graph fibrations, input trees and fibers.}

A \emph{graph fibration} is a graph morphism $\varphi: G \rightarrow B$
between a graph $G$ (the total space) and a graph $B$ (the
base), in which, for every (pre-image) node $i$ in $G$, and every (image) edge $e'$ in $B$
targeting the image of $i$ (i.e.  $i'=\varphi(i)=t(e')$), there is a \emph{unique}
edge $e$ (in $G$) targeting $i$ ($i=t(e)$) whose image is
$e'=\varphi(e)$. Here, $t(e)$ denotes the \emph{target} node of edge $e$, for more mathematically rigurous definitions we refer to the SI Section~\ref{sec:SI:fibrations}.

\begin{figure*}[!ht]
    \centering
    \includegraphics[width=.8\textwidth]{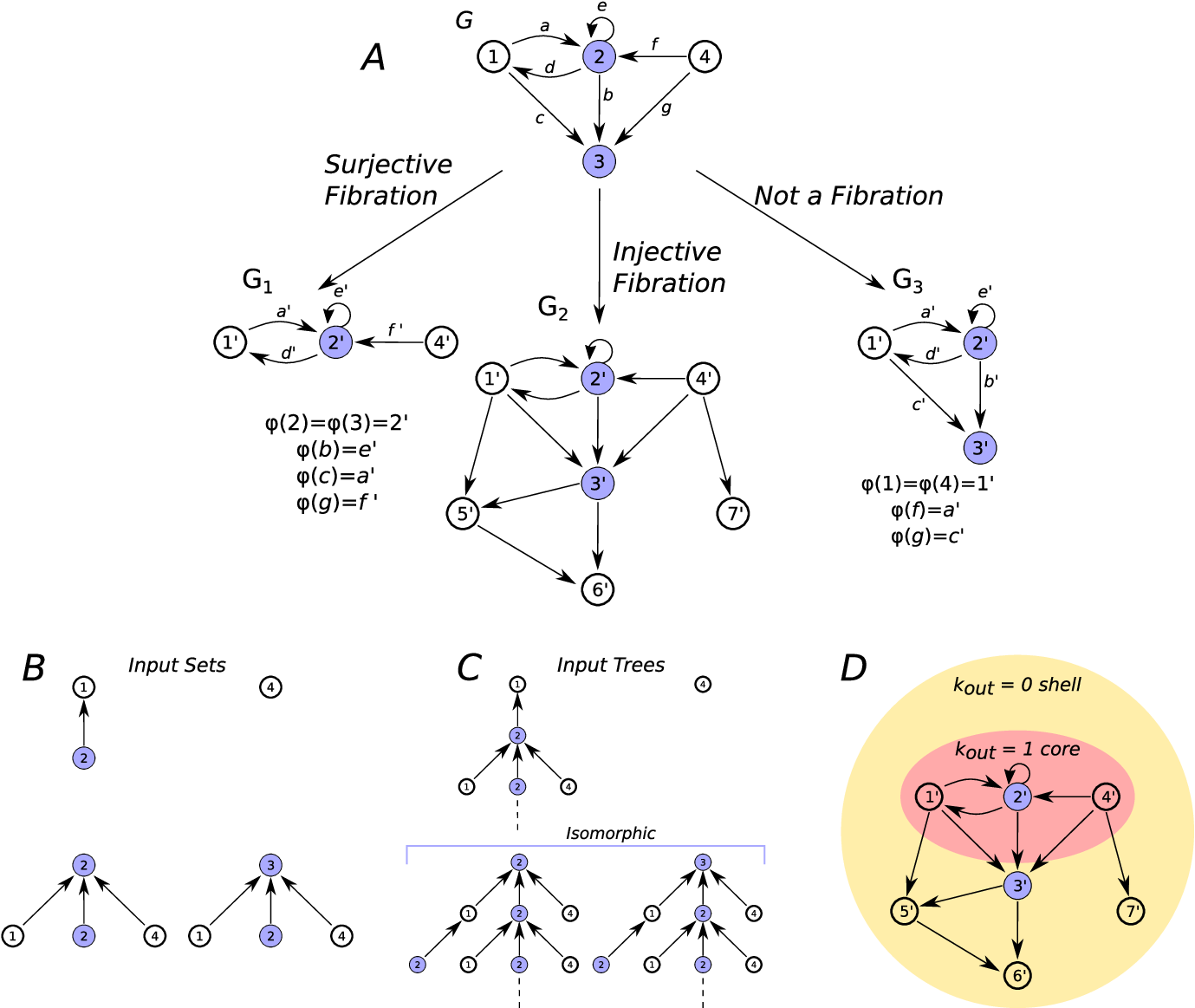}
    \caption{{\bf Fibration and $k_{out}$-core decomposition.} \textbf{A} Graph $G$, a subgraph of the GRN of \emph{E.~coli}, shows a Fibonacci
      building block with class number $|\varphi = 1.6180, \ell = 2\rangle$~\cite{morone2020fibration, 2020predict}. All three mappings are morphisms since the images of all the nodes in $G_1, G_2$ and $G_3$ are connected only when corresponding nodes in $G$ are connected, respecting the incidences. The mapping $G \rightarrow G_1$, in the left, corresponds to a surjective fibration: all nodes with isomorphic input trees are collapsed to one (nodes $2$ and $3$ collapsed to $2'$), all input trees are preserved, hence the lifting property is satisfied. Mapping $G \rightarrow G_2$ is an injective fibration. Indeed, it is easy to see that the original graph is embedded in $G_2$ making this map a morphism where all input trees are preserved. Some nodes and edges are added but without breaking the original input trees. 
      The mapping $G \rightarrow G_3$, which maps node $4$ to $1'$ does not correspond to a fibration given that the input-tree of node $4$ (seen on \textbf{B}) is not preserved in its image node $1'$ in graph $G_3$, the same problem occurs with the images of nodes $2$
      and $3$ ($2'$ and $3'$ respectively), their input trees are not preserved as the former input from node $4$ is lost. Edges $a'$ and $c'$ cannot be uniquely lifted at $\varphi(2)$, since they need to be lifted to $a, f$ and $c, g$, respectively, for the mapping to be a morphism. 
      In practical terms, since the input from node $4$ is lost, graph $G_3$ represents an entirely different dynamical system from graph $G$. If the graph $G$ represents a GRN, genes $2'$ and $3'$ in $G_3$ would have a different expression pattern than genes $2$ and $3$.
      \textbf{B} Shows the input sets and \textbf{C} the input trees of nodes in graph $G$. The input set of node 2 is repeatedly attached to node 2 in every layer of the trees, due to its self-loop, this process is repeated ad infinitum. As a result, the input trees of nodes 1, 2 and 3 are infinite; however, since $G$ has only 4 nodes, it suffices to verify the isomorphism up to the third layer of their trees, hence nodes 2 and 3 are determined to have isomorphic input trees. \textbf{E} Example of the $k$-core decomposition of graph $G_2$ from \textbf{A}. Even though node $5'$ on the outer $k_{\rm out} = 0$ shell (in red) does have
      one output, once nodes $6'$ and $7'$ in the shell are removed,
      it will then be left with no output and will be removed as
      well. All the remaining nodes in the $k_{\rm out} = 1$ core have at
      least $1$ output after doing this process.}
    \label{fig:fibration}
\end{figure*}

Simply put, every edge targeting an image node $i'=\varphi(i)$ can be \emph{uniquely} lifted to an
edge in $G$ targeting its pre-image $i$. This condition is called the lifting
property~\cite{boldi2002fibrations}. Crucially, this means that the inputs of any node are preserved in the base graph.
In Fig.\ref{fig:fibration}A we illustrate the definition of graph fibration, we see examples of three different morphisms, one is not a fibration while two of them are: a surjective fibration, and an injective fibration.

The question now then becomes how to indentify the synchronous clusters from the network structure. To do this we have to look at the input history of each node via their \emph{input tree}. Nodes with identical input histories will become synchronous since they receive the exact same history of signals.

The {\it input tree} $T_{i}$, for a node $i$ in graph $G$, is made up of the set of all pathways in $G$ ending at node $i$~\cite{morone2020fibration}. In order to construct the input tree of a node, we start by constructing its \emph{input set:} the set of incoming edges to the node in question, along with their respective source nodes (see Fig.~\ref{fig:fibration}B). We then attach to this rooted tree, the input sets of the incoming nodes, and so-on recursively, to obtain the input tree.
Therefore, the input tree of a given node summarizes the regulatory pathways of the
network that reach this node. This allows us to group and classify the nodes based on the \emph{'history'} of the signals they receive. Fig.~\ref{fig:fibration}C shows the input trees for the nodes in graph $G$.

Upon determining the sets of nodes with \emph{isomorphic input trees}, we determine the \emph{fibers}: sets of nodes with {\it identical} input history. Fibers are called {\it balance colorings} or {\it equivalent relations} in other branches of mathematics, dynamical systems and chaos~\cite{stewart2003symmetry,golubitsky2006nonlinear}. For infinite input trees it suffices to show the isomorphism up to $N_G - 1$ layers of the trees (where $N_G$ is the number of nodes in $G$) to determine the isomorphism~\cite{norris1995universal}. 
The nodes in a fiber receive the exact same
signaling tree, which makes the paths of the signals rooted at
them redundant, revealing a sense of symmetry. This in turn makes the nodes in the same fiber
symmetric in terms of signal processing or information flows in the
network.

Having the set of fibers, we can construct the fiber building blocks of the network, Fig.~\ref{fig:circuitsOverview} shows their canonical forms. These are constructed by obtaining the induced graphs of each fiber along with their regulators (also including the nodes along the path(s) forming a feedback loop from the fiber to the regulators, if any).

{
The notion of fibrations can be extended to graphs with various types
of edge, such as a GRN, which can have edges
corresponding to several types of interaction: activation, inhibition, other types of interaction. 
In this case, for two input trees to be isomorphic not only do they need to have the same topology, but the type of edge must be the same.} 

In the case of gene regulation networks, fibers are sets of genes that are coexpressed in their activity; this is the case for the blue nodes in Fig.~\ref{fig:fibration}A. 

\subsection{Complexity Reduction by Symmetries (ComSym) method step by step}


We developed a stepwise method called "Complexity Reduction by Symmetries" (ComSym) to reduce any signaling network to its computational core. The method aims to clarify the structure of the network, help to understand the decision-making processes performed by the network. ComSym can be applied to any directed network, even outside biology, in which edges represent a signal transmitted from one node to another and provides insights about the collective dynamics within the network based solely on its topology. Regardless of the exact model, it can be used to describe the underlying dynamical system 
since the method does not depend on the actual form of the admissible ODEs in the graph~\cite{golubitsky2006nonlinear, stewart2003symmetry, boldi2002fibrations, deville2013modular}.

The ComSym method for network reduction consists of five steps. A more detailed description is given in  SI section \ref{sec:SI:ComSymStepByStep}.

\paragraph{Steps I and II: Reducing a signaling network to its computational core}
Step I, \emph{'collapsing'}, removes the symmetries in signaling flow in the network that originate from fibration
symmetries. This step is based on Lemma 5.1.1 from
Ref.~\cite{deville2013modular} which proves that the dynamics of a
network is preserved when all symmetries are eliminated by a
surjective graph fibration. 

In a graph fibration, multiple edges
targeting the same node in $G$, cannot be collapsed to fewer edges in
$B$, nor can new edges targeting the image of a node be
added. Crucially, though, nodes \emph{can} be collapsed if they belong to the same fiber
(given their redundant, or symmetric, input trees). 
This is the crucial feature of graph fibrations, once we have identified the redundant pathways, we can eliminate the redundancies without losing any "information pathways"; only the redundancy is removed while the dynamics are preserved. This is done by collapsing the fibers into a single representative node for each fiber. Such a surjective fibration is shown on Fig.~\ref{fig:fibration}A.





Step II, \emph{'pruning'} the loose ends of the network, makes use of the direction of the signal flow in the network. 
An injective fibration (also shown in Fig.~\ref{fig:fibration}A) formalizes the idea that under certain conditions, a subset of the constitutive elements of a system may drive the dynamics of the entire system.
Lemma 5.2.1 in Ref.~\cite{deville2013modular} shows that the dynamics of the nodes in the "outer" layer is driven by the dynamics of the (inner) core network. Therefore, the dynamical behavior of the "core" network can be studied separately and further used to scrutinize the dynamics of the "outer" layer. 

Reducing the network to its core is performed by applying an inverse injective fibration: the $k$-core decomposition of the network, which identifies an "outer" layer (shell or periphery) of nodes that do not send signals to the "inner" core of the network, Fig.~\ref{fig:fibration}D shows this decomposition.
This step can be thought of as trimming the loose ends of a tree. The removed nodes belong to the "null"-paths, the dead-end paths in the network: the nodes that do not posses any output, along with the nodes that exclusively regulate them, and so on iteratively. These are the nodes that send signals to the peripheries of the network and other parts of the bacteria, such as the metabolic network via genes that express enzymes for metabolic functions, such as sugar consumption, among others. For a more detailed and rigorous explanation see SI Section~\ref{sec:inj}.

\paragraph{Step III: Large-scale structure of the Minimal network: Strongly Connected Components and signal vortices}

The previous two reduction steps yield the core subnetwork that controls the
dynamics of the entire system, the minimal network. 
After pruning the loose ends of the network, since all the dead-end paths are lost, all the remaining paths self-cross at some point. 
The minimal network will therefore always consist of the SCCs and the nodes regulating them (assuming that there are SCCs; for an acyclic network no minimal network would be obtained). For this reason, we want to understand how the network decomposes into SCCs, the large-scale structure of the minimal network.
Here, SCCs represent the smallest computational subunits that cannot be further reduced neither by the fibration symmetries nor by smaller strongly connected components. 

Hence, on Step III, we decomposed this minimal network into its SCCs. The nodes that do not belong to the SCCs are connectors between them or controllers (external regulators) nodes that send signals to the SCCs but do not receive any signals back from them (otherwise they will be part of the SCC by definition). 
As a consequence, the most trivial result that could be expected is that the core of any network corresponds to only one single SCC, i.e. virtually no structure in the core network. This is in fact the case for most randomized (degree-preserving) versions of these networks as will be shown later in more detail (Section~\ref{sec:stat sig}), but not at all the case for the GRNs studied.

\paragraph{ Steps IV and V: Small-scale structure inside the SCCs: logic circuits and cycles}

The last two steps consist of understanding the small-scale structures within the SCCs: the logic circuits (Step IV) and the bigger cycles connecting them (Step V).

Since the inception of
synthetic biology and the first genetic circuits designed two decades
ago~\cite{gardner2000construction, elowitz2000synthetic, stricker2008fast, cameron2014brief, khalil2010synthetic}, it is known that simple
genetic circuits can perform the basic logic operations
necessary for any computational device, such as memory storage and
timekeeping~\cite{elowitz2000synthetic, stricker2008fast, tyson2003sniffers, leon2016computational, dalchau2018computing}. Such circuits are
constructed using feedback loops (and hence they will always be embedded in the SCCs), both positive and
negative~\cite{tyson2003sniffers, dalchau2018computing}, and are
executed by synthetic switches and oscillators designed from simple
components such as interacting genes or protein-protein interactions~\cite{tyson2003sniffers, dalchau2018computing}.

The most basic memory circuit corresponds to the toggle-switch~\cite{gardner2000construction}, analogous to a bistable flip-flop in electronics~\cite{tanenbaum2016structured}
that stores one bit of information given that it has
two possible stable and reciprocal states.
It is comprised by two mutually repressive (MR) genes with different inputs for each gene, the \emph{'set' (S)} and \emph{'reset' (R)} switches. While for time-keeping, oscillating circuits can be obtained by a "frustrated" signaling chain, most commonly by a negative feedback loop (NFBL) driving the system back
and forth between the stable steady states~\cite{tyson2003sniffers}. The most simple form of this is two genes with a NFBL between them, where one gene activates the other, while at the same time it is being inhibited by the other. The presence and type of self-regulations in these cases changes the specific dynamics: no self-regulations requires noise to drive the oscillations~\cite{geva2010fourier, lahav2004dynamics}, while the most robust version corresponds to the Smollen oscillator~\cite{stricker2008fast}. Other forms of oscillating circuits are also possible~\cite{elowitz2000synthetic, leon2016computational, dalchau2018computing, leifer2020circuits}.
These circuits have been known from before, but here we are able to reinterpret them as broken symmetry versions of symmetric fiber building blocks, see SI Section~\ref{sec:circuits hierarchy} for further explanation.

In Step IV we
systematically look for circuits capable of logic computations
in the minimal network. Since these circuits can be
artificially constructed to perform computations, it is reasonable to
expect to observe them, or some close variation, at the core
computational subset of the network. We would expect to find both memory
storage circuits as well as oscillating circuits for timekeeping. In
the case of the GRN of simple model bacteria, we would expect to
observe the simpler forms of these known genetic circuits from
synthetic biology. 
Indeed, we find the presence of circuits closely resembling all
these circuits in the minimal network driving the GRN. This suggests that we can 
understand this minimal network as a logical computational machine. 

Furthermore, in Step V, we study the structure of the signal vortices, the SCCs, that make up the large-scale structure of the minimal GRN and 
the interconnectedness between the different logic circuits present. An SCC is composed of a complicated arrangement of feedback loops between its constituent nodes, as such we probe its structure by studying the independent simple cycles present in the minimal GRN. These cycles in themselves represent a form of longer-term memory, responsible for the interconnectedness of the logic circuits, where signals loop between different logic circuits. 

Crucial to the dynamics of these circuits are feedback loops between
different genes. This implies that the circuits are always embedded
in the SCCs of the network. Given that the SCCs are preserved after
our reduction method, we know that we are not losing any logic
components of the network as we reduce it. This means that we can interpret
the SCCs of the network as the modules where the logic
computations are performed.


\section{Results}

\subsection{Application of ComSym to bacterial gene regulatory networks}



\begin{figure}[t!] 
    \centering
    \includegraphics[width=.675\linewidth]{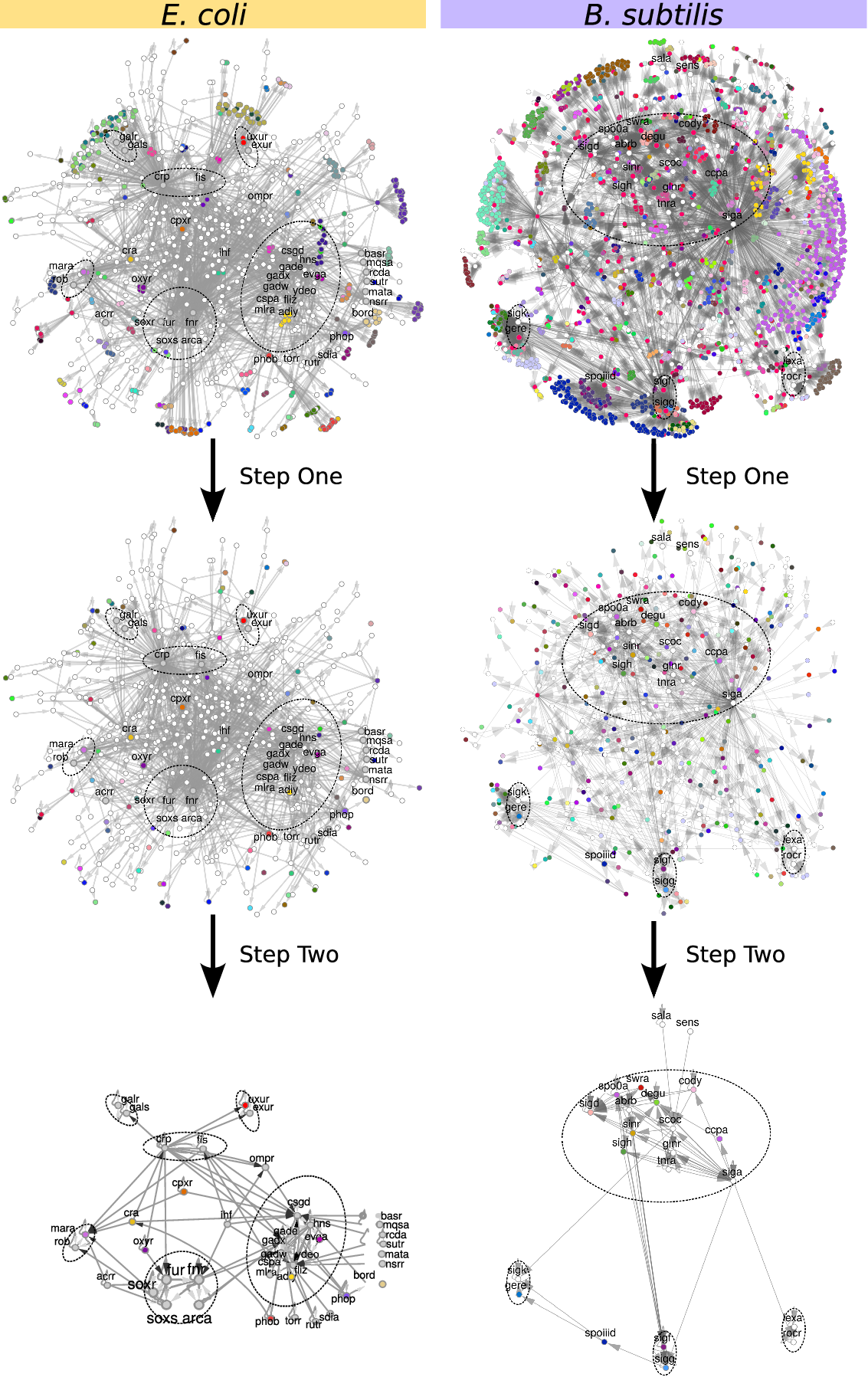}
    \caption{{\bf Network reduction of the GRN of \emph{E.~coli} and \emph{B.~subtilis}}. For \emph {E.~coli}'s network, we start with the network from Fig.~\ref{fig:main}, rearranged to show the outward flow of signals from the minimal network and with genes in the same fibers colored the same. SCCs are enclosed by ellipses. Genes names are shown only for genes that are part of the minimal network. Most of the genes belonging to fibers can be seen located in the periphery (outer regions) of the network. 
    \textit{Step one} of the ComSym procedure collapses all fibers into one representative node, resulting in the base network obtained from the minimal surjective fibration.
    \textit{Step two} uses a $k$-core decomposition to removes all the dead-end paths ending at nodes with no output, resulting in the minimal network, with only 42 nodes for \emph{E. coli} and 22 in \emph{B. subtilis}. 
    Both minimal networks have a master SCC that regulates the rest, connector nodes connecting different SCCs as well as controller nodes sending inputs to the SCCs.}
    \label{fig:steps}
\end{figure}

To demonstrate the use of our method, we applied it to two of the most widely studied bacterial GRNs, the networks of \emph{E.~coli}~\cite{shen2002network, milo2002network,   thieffry1998specific, martinez2008functional} and \emph{B.~subtilis}~\cite{gbmn:08, fotc:16}. Their step-wise reduction is shown in Fig.~\ref{fig:steps}, while some statistics are given in Tables~\ref{tab:reduction},~\ref{tab:breakdown}, and~\ref{tab:circuits}.


We observe a rich structure of connections between multiple SCCs, both direct connections and through longer pathways crossing \emph{bridge} or \emph{connector} nodes. This suggests that the structures we observed are not the result of randomly generated networks, based on the fact that 
the number of SCCs for each GRN is more than 5 standard deviations away (see Z-scores in SI Table.~\ref{tab:rand}) from the average number of SCCs observerd in the randomized degree-preserved generated networks (more on Section~\ref{sec:stat sig}), indicating it is extremely unlikely for these structures to originate out of pure chance.

In both bacteria studied, we obtained a rich but transparent structure: the SCCs of the network receive inputs from outside controlling nodes as well as some connecting nodes between the different SCCs, responsible for transmitting signals between the SCCs. This also yields a simple and modular interpretation of the minimal GRN as the computational core driving the dynamics of the entire GRN.

In both cases, there is a \emph{master SCC} regulating the other SCCs.  The general flow of information can be described like this: external signals enter the SCCs, through the set of controller genes (regulatory genes participating in one-, two-, or three-regulations of SCCs), where they are fed to the logic circuits, logic computations occur within the SCCs, and the signals then emanate outward from the SCCs and from the \emph{master} SCC to the other SCCs. The output of the minimal network is then propagated outward to the fibers (clusters of co-expressed genes) regulated by the SCCs in the periphery of the network (see Fig.~\ref{fig:surjmin}) and to other parts of the cellular network, such as the metabolism. Thus, the SCCs act as decision-making units that activate the fibers under their control.

Together, the method lead to the identification of the
function for every single gene in the minimal GRN as belonging to three
general classes of genes:

\begin{itemize}
\item (1) A set of synchronized symmetric fibers
\item (2) Regulators of the SCCs
\item (3) logic circuits within the SCCs (often arising from an $n=2$ fiber symmetry breaking), which can be further classified into:
    \begin{itemize}
    \item (3.1) memory devices (toggle switches)
    \item (3.2) oscillators.  
    \end{itemize}
\end{itemize}

Perhaps it is import to note that this result does not depend on the simplification assumptions or even the specific ODEs used to model the genetic interactions, this is purely the network structure. The specific details of the parameters affect how large the synchronization within the fibers is, while the choice of function used determines the specific activation levels, times, and bifurcation processes that impact the specific details of the internal computations of the logic circuits; however, the overall flow and decomposition of the network is a result of network structure alone. 

The code implementing the method for reproducing the entire analysis in the present paper can be found in \url{https://github.com/luisalvarez96/MinimalTRN} and a more detailed explanation as well as a pseudocode can be find in SI Section~\ref{SI:Algorithm}.




\begin{table}[!t]
  \centering
  \begin{tabular}{lrrrr}
    & \multicolumn{2}{l}{\textbf{\cellcolor{myEcoliColor} \emph{E.~coli}}} & 
    \multicolumn{2}{c}{\textbf{\cellcolor{myBsubColor} \emph{B.~subtilis}}}\\
    \toprule
    \textbf{Reduction step}     & \textbf{\emph{Genes}} & \textbf{\emph{\%}} & \textbf{\emph{Genes}} & \textbf{\emph{\%}}\\
    \midrule
    Step 0.0: Full Genome & 4,690 & -- & 6121 & --\\
    \hline
    Step 0.1: GRN (non isolated TFs)   & 1,843 & 100\% & 2482 & 100\%\\
    Step 0.2: operon-collapsed GRN    & 879 & 48\% & -- & --\\
    \hline
    Step 1: Base-GRN (collapsing fibers) & 555 & 30\% & 521 & 21\%\\
    Step 2: Minimal GRN (after pruning) & 42 & 2\% & 22 & 0.9\%\\
    \bottomrule
  \end{tabular}
  \caption{\textbf{Gene counts of original and reduced GRNs.} The full genomes for \emph{E. coli} and \emph{B. subtilis} contain 4,690 genes (according to RegulonDB
    \cite{tierrafria2022regulondb}) and 6121 genes(obtained from SubtiWiki~\cite{SubtiWiki}), respectively.
    Among all genes only 1843 and 2482, respectively, express TFs with known interactions. The first reduction step in \emph{E.~coli} was performed by a trivial fibration (collapsing the operons, which are trivial fibers), before applying ComSym. This could also be seen as a part of Step 1 and as such is not needed for \emph{B.~subtilis}. 
    }
  \label{tab:reduction}
\end{table}

\begin{table}[!h]
  \centering
  \begin{tabular}{lrrrr}
    & \multicolumn{2}{l}{\textbf{\cellcolor{myEcoliColor} \emph{E.~coli}}} & 
    \multicolumn{2}{c}{\textbf{\cellcolor{myBsubColor} \emph{B.~subtilis}}}\\
    \toprule
    \textbf{GRN breakdown}     & \textbf{\emph{Genes}} & \textbf{\emph{\%}} & \textbf{\emph{Genes}} & \textbf{\emph{\%}}\\
    \midrule
    GRN  & 879 & 100\% & 2482 & 100\%\\
    Nodes in fibers    & 416 & 47.3\% & 2263 & 91.2\%\\
    \hline
    $k_{\rm out}$ shell (fiber-collapsed nodes) & 513(82) & 58.4\% & 499(290) & 20.1\%\\
    Nodes in SCCs (fiber-collapsed nodes) & 24(4) & 2.73\% & 19(11) & 0.77\%\\
    Connectors (fiber-collapsed nodes) & 18(6) & 2.05\% & 3(1) & 0.12\%\\
    \bottomrule
  \end{tabular}
  \caption{\textbf{Statistics and fiber coverage of the two GRNs.} For \emph{E.~coli} we start with the 879 operon-GRN from Step 0.2 (see previous Table~\ref{tab:reduction}). For \emph{E.~coli}, Step 1 collapses the 416 nodes within fibers into 92 fiber-collapsed nodes (one for each fiber), to give the collapsed-fibers 555 nodes Base-GRN. For \emph{B.~subtilis}, the 2263 fibered nodes are collapsed into 302 fiber-collapsed nodes (one for each fiber), resulting in the 521 nodes Base-GRN.
    Step 2 removes all the nodes in the outer shell, including 82 of the fiber-collapsed nodes for \emph{E.~coli} and 290 for \emph{B.~subtilis}, thus leaving only the minimal GRN. 
    The minimal networks are composed of the nodes in SCC and the connectors nodes.
    The $k$-shells are actually much bigger, but we count the genes inside them after collapsing the fibers, resulting in only 82 fiber-collapsed nodes (in the case of \emph{E.~coli}) instead of counting all the original nodes that belong in these fibers.}
  \label{tab:breakdown}
\end{table}

\begin{table*}[t!]
\small\sf\centering
  \begin{tabular}{lll}
    \bf Circuit type & \cellcolor{myEcoliColor} \bf \emph{E.~coli}  & \cellcolor{myBsubColor} \bf \emph{B.~subtilis} \\
    \toprule
    \bf Toggle-switch type & \emph{galS-galR}, \emph{uxuR-exuR}, & \emph{lexa-rocr}, \emph{glnr-tnra}\\
    & \emph{csgD-fliz} \\
    \hline
    \bf Oscillator type & \emph{rob $\mapsto$ marA}, \emph{soxS $\mapsto$ fur}, & \emph{sigk $\mapsto$ gere}, \\
    & \emph{cspA $\mapsto$ hns}, \emph{gadX $\mapsto$ hns}, & \emph{siga $\mapsto$ spo0a} \\
    \hline
    \bf Lock-on types & & \emph{sigf-sigg}, \emph{siga-sigh},\\ & & \emph{siga-sigd}, \emph{sigd-swra} \\
    \hline
    \bf Capable of various types & \emph{crp-fis}, \emph{gadW-gadX},  \\
     & \emph{fnr-arcA} \\
    \hline
    \bf FFF type & \emph{ihf}$\mapsto$ \{\emph{fis,fliz}\} $\mapsto$ \emph{gadX-gadE} \\
     & \emph{ihf}$\mapsto$ \{\emph{fnr,fliz}\} $\mapsto$ \emph{gadX-gadE} \\
     & \emph{ihf}$\mapsto$ \{\emph{fnr,fliz}\} $\mapsto$ \emph{gadW-gadE} \\
    \bottomrule
  \end{tabular}
  \caption{\textbf{List of gene circuits in \emph{E.~coli} and \emph{B.~subtilis}.} The circuits found in \emph{E.~coli} are described in detail in  Fig.~\ref{fig:all_circuits} and discussed at length in SI Section~\ref{sec:circuits hierarchy}}
  \label{tab:circuits}
\end{table*}

\subsection{Minimal gene regulatory network of \emph{Escherichia coli}}
\label{sec:ecoli}

The gene regulatory network of \emph{Escherichia coli}, as provided by RegulonDB, contains around 4,690 genes; however, the majority of these genes code for proteins that do not regulate any other genes, but are enzymes, structural proteins, etc., or they may also constitute TFs whose interactions are not yet well known. Since these genes do not have a clear regulatory role within the GRN, we do not consider them in our analysis, and keep only the annotated transcription factors. This leaves us with 1,843 genes involved in the GRN. 

\paragraph{Network reconstruction and reduction to Transcriptional Units (TUs)} 
Operons are clusters of contiguous genes that get transcribed together by the same polymerase as a unit, hence being "trivially" synchronized. Some operons however, have internal promoter regions that allow some of the genes of the operon to be transcribed in different transcription units (TUs) without the need to transcribe the full set of genes in the operon.
To simplify our analysis, we also collapse transcriptional units (genes in the same operon, under the control of the same promoters) operons into single nodes, because such genes would form \emph{"trivial"} fibers. To do so, operons with such internal promoters are split into its different transcription units (TUs) in our network, which are now treated as gene nodes. This trivial reduction, which is in reality a part of our step 1 of collapsing, shrinks our initial 1843-genes GRN to 879 nodes. Strictly speaking, this reduction is part of our Step 1 and can be done together in only one step, however, we took this step to avoid looking at operon's trivial fibers, during initial analysis (as was done in Ref. \cite{morone2020fibration}). 

\paragraph{Network reduction}
Applying the symmetry fibration to \emph{Escherichia coli}'s GRN results in just 555 nodes, 30\% the size of the original network, as seen in Table~\ref{tab:reduction}. Step 1 could also have been applied directly to reduce the initial 1843 network to the 555-node base network. 
Most of the genes belonging to fibers are located in the periphery of the network. 
Various nodes with notably high out-degree in the network don't belong to the core network. After this, the remaining 555 nodes are reduced by taking the $k_{out}=1$-core of the network into just only 42 nodes. These 42 nodes make up the minimal GRN which corresopnds to the computational core of the GRN. 

\paragraph{Large-scale structure of \emph{E. coli}'s minimal GRN}
\label{sec:ecoli2} The structure of the core network (as shown in Fig.~\ref{fig:main} on the right), is obtained from the reduction process illustrated in Fig.~\ref{fig:steps}. We find that {\it E.~coli}'s minimal GRN is composed of 6 SCCs, see Fig.~\ref{fig:steps}. 

\begin{figure*}[t!]
    \centering
    \includegraphics[width=.7\linewidth]{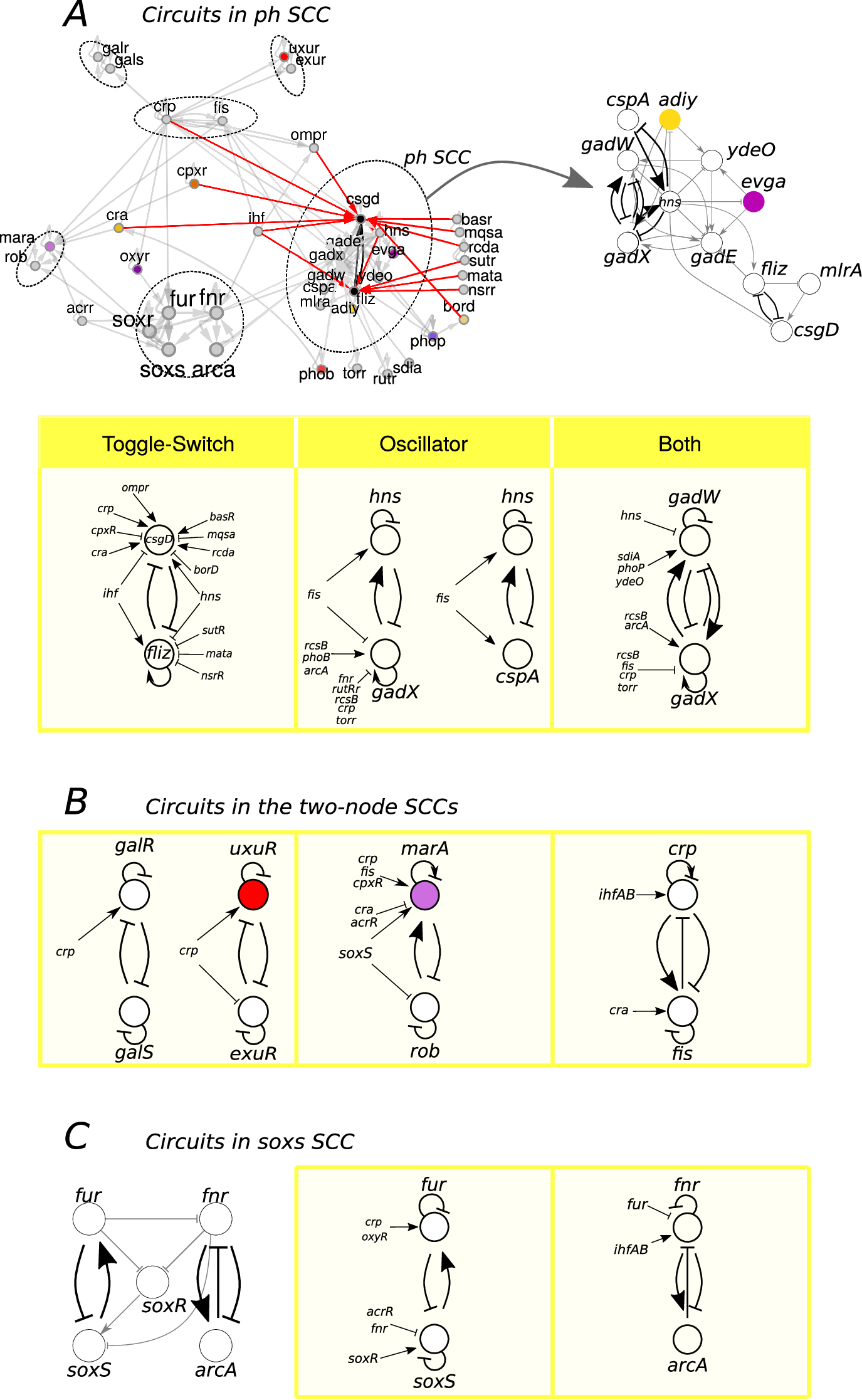}
    \caption{{\bf Circuits in the GRN of \emph{E.~coli}}. \textbf{A} The minimal GRN of \emph{E.~coli}  and the
      circuits embedded in it, shown
      with red links for the symmetry breaking inputs to the
      toggle-switch \emph{fliz-csgd}. The biggest SCC is in charge mostly of pH responses. Colored nodes represent fibers. \textbf{B} The two-node SCCs and
      \textbf{C} the \emph{soxS} SCC and its circuits. 
      For each circuit, the incoming signals that
      break the symmetry are shown. }
    \label{fig:all_circuits}
\end{figure*}

The central subunit of this minimal GRN is the carbon SCC: \emph{crp-fis}. 
Which serves as the carbon utilization subnetwork \cite{tierrafria2022regulondb} controlling a set of TF and enzymes involved in the catabolism of the different sugars and thus is one of the main components for the life of the cell.
Another SCC, with 5 nodes, involved in responses to \emph{stress}~\cite{tierrafria2022regulondb} in the cell. We call this the \emph{soxS} SCC, made up of \emph{fur-fnr-arca-soxs-soxr}, bottom center of the minimal GRN in Figs~\ref{fig:steps} and~\ref{fig:all_circuits}.
Additinally, we obtain one with 11 nodes which mostly regulates the cell's \emph{pH} response~\cite{tierrafria2022regulondb} (we will refer to it as the \emph{pH} SCC, lower right corner of the minimal GRN in Figs~\ref{fig:steps} and~\ref{fig:all_circuits})
The remaining three SCC are: \emph{marA-rob} SCC, which controls a number of genes involved in resistance to antibiotics~\cite{tierrafria2022regulondb}; the \emph{uxuR-exuR} SCC, which are involved in regulation of the transport and catabolism of galacturonate and glucuronate \cite{tierrafria2022regulondb}; and the \emph{galR-galS} SCC related to the import and catabolism of galactose~\cite{tierrafria2022regulondb}. For a detailed description of the signal vortices, see SI section \ref{sec:SI.DetailsOnGRNs}.

The three main SCCs are connected in a forward manner: carbon~$\rightarrow$~ph, carbon~$\rightarrow$~stress, and stress~$\rightarrow$~ph; thus forming a feedforward loop representing the core of the genetic computing system. The same structure is observerd between the carbon SCC, the stress SCC and \emph{marA-rob} SCC (carbon~$\rightarrow$~\emph{marA-rob}, carbon~$\rightarrow$~stress; stress~$\rightarrow$~\emph{marA-rob}). The two remaining SCCs only receive information from the carbon SCC.


\paragraph{Logic circuits present in \emph{E. coli}'s minimal GRN} Inside the SCCs of the GRN we found various genetic circuits that resemble logic circuits designed and implemented in the synthetic biology literature. In total, 12 different pairs of genes were found to be involved in a number of logic circuits. Fig.~\ref{fig:all_circuits} shows all the circuits found in \emph{E. coli}'s GRN, as well as all the inputs to them that break their symmetry and drive their computing. For a detailed description of the gene circuits and what the literature tells us about their possible dynamics, see SI section \ref{sec:SI.DetailsOnGRNs}.

\begin{figure}[t!] 
    \centering
    \includegraphics[width=.7\linewidth]{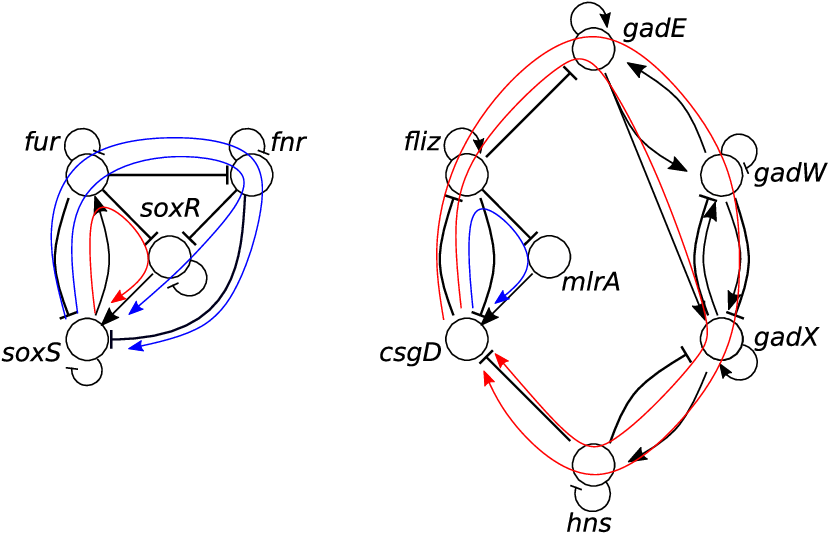}
    \caption{\textbf{Some simple directed cycles in
      \emph{E.~coli}.} The networks shown are  different cycles that cross through the
      logic circuits \emph{soxS}-\emph{fur} (left) and \emph{csgD-fliz} (right). Arrow colors denote the overall sign (overall activation: blue; overall inhibition: red).}
    \label{fig:cycles}
\end{figure}

\paragraph{Simple directed cycles in \emph{E. coli}'s minimal network} In total we found 41 simple directed cycles in the minimal core of \emph{E.~coli}'s GRN. Four of them are the two-node SCCs, 5 are located in the \emph{soxS} SCC and the remaining 32 are located in the \emph{ph}
SCC. 
Each of the cycles contains at least a pair of nodes that between them form a logic circuit. This means that each cycle longer than 2 nodes passes through at least two nodes that are also connected by a logic circuit (all logic circuits themselves are, of course, two-node cycles).
For example, in the case of \emph{soxS} SCC we observe 2 two-node cycles (circuits \emph{soxS~$\mapsto$~fur} and \emph{fnr-arcA}, see Fig.~\ref{fig:all_circuits}C), while the remaining 3 cycles in this component all pass through \emph{soxS~$\mapsto$~fur} and as such can be considered longer loops
from \emph{soxS} to \emph{fur}, this is shown in
Fig.~\ref{fig:cycles} on the left. In many case, the cycles
even pass through multiple nodes that are connected by a logic
circuit to each other. This is also visible on the right of Fig.~\ref{fig:cycles}, where the longer loops passthrough \emph{gadE-gadW},
\emph{gadW-gadX}, and \emph{gadX-hns} all of which are different logic circuits by themselves. All of these the loops illustrated can be
considered loops of various lengths between the circuits \emph{soxS
  $\mapsto$ fur} and \emph{csgD$-$fliz}. A complete list of all cycles
is available in the provided repositories.


  
\subsection{Minimal gene regulatory network of \emph{Bacillus subtilis}}
\label{sec:bac}

Next, we analysed the GRN of a second example organism, the soil bacterium \emph{Bacillus subtilis}. \emph{B.~subtilis} is a main model organism for Gram-positive bacteria with a well-studied GRN. The data used for
\emph{B.~subtilis} included not only transcriptional activators and
repressors, but also sigma factors, which play a bigger role in this organism~\cite{gbmn:08}. A sigma factor binds to the
promoter region of a target gene to enable its transcription, and
different sigma factors target specific groups of genes, for
example, genes involved in stress responses. In our analysis, we
treated sigma factors as inducers, just like activating specific TFs. Previously, the integrated metabolic and regulatory network of \emph{B.~subtilis} has been broken down into functional modules, locally regulated or regulated by a global regulator~\cite{gbmn:08, fotc:16}. However, the overall structure of this network, how the modules interact with each other, and the overall flow of information or signals between these are still not entirely understood. Similarly as in~\emph{E.~coli}, our method revealed a structure of the gene regulatory network for this bacteria and to identify the possible logic circuits at the core of the network. Like for \emph{E.~coli}, all nodes in the GRN could be classified:
in the original GRN, each node either belongs to a fiber, belongs to a
logic circuit, or sends inputs to circuits.

\begin{figure}[!ht] 
    \centering
    \includegraphics[width=0.45\linewidth]{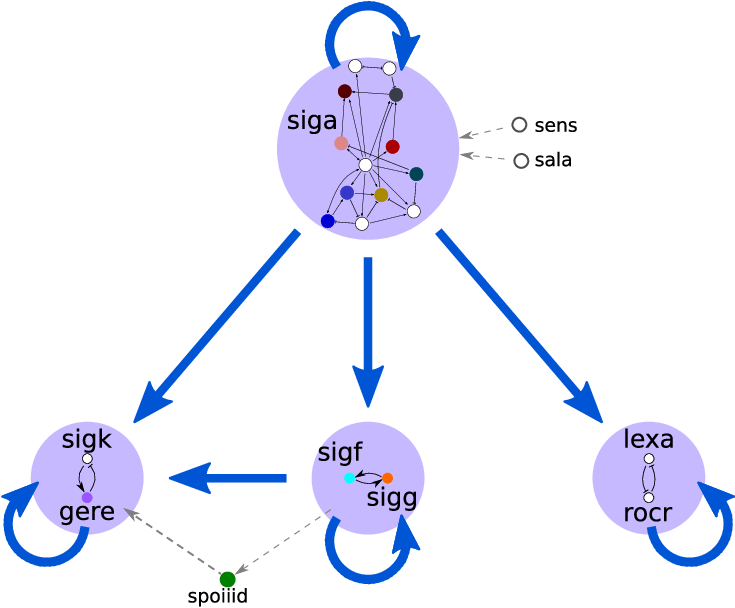}
    \caption{\textbf{Sketch of reduced GRN of \emph{B.~subtilis}}. The 4 SCCs
      are shown: \emph{siga} SCC, \emph{sigk-gere} SCC, \emph{sigf-sigg} SCC, and \emph{lexa-rocr} SCC. The signaling flows between them: with the \emph{siga} SCC the hub controlling the other three SCC and being fed information signals by the two controllers \emph{sens} and \emph{sala}, controller node \emph{spoiiid} connects the \emph{sigk-gere} SCC and \emph{sigf-sigg}. A feedforward structure
      between the \emph{siga} SCC and the \emph{sigk-gere} SCC is visible.}
    \label{fig:bac}
\end{figure}

\paragraph{Network reduction}
The first reduction by fibration reduces the network size to $21\%$ of its original size. The $k$-core reduction led to a further reduction to just $0.9\%$ of the
original nodes, as can be seen in Table~\ref{tab:reduction}. The resulting
core GRN for {\it B.~subtilis} is shown in the right column of Fig.~\ref{fig:steps} and corresponds to just 22 nodes.


\paragraph{Large-scale structure of \emph{B. subtilis}'s minimal GRN}
The structure of the minimal network is shown in Fig.~\ref{fig:bac}.
Like in {\it E.~coli}, the resulting minimal set of TFs obtained for
{\it B.~subtilis} contains only the SCCs, the fibers that connect them
and the nodes that send signals to control them. However, it is particularly interesting how this minimal gene regulatory network is
smaller, with just 22 nodes in 4 SCCs, than the one for {\it E.~coli},
given that the original network is larger (2482 nodes). In contrast to
{\it E.~coli}, \emph{B.~subtilis}' minimal GRN is almost exclusively
the SCCs with only 3 controlling nodes whereas in {\it E.~coli}
there are 18 controlling nodes that inform the SCCs modules.

The structure of the minimal computational core in \emph{B.~subtilis}
is simpler than in \emph{E.~coli}. It consists of only 4
SCCs: the \emph{siga} SCC, a large central SCC as a hub composed of 13 nodes, which
regulates the other three SCCs, each containing only two nodes. This SCC hub is controlled by two control nodes (\emph{sala} and \emph{sens}). Importantly, we also find
a feedforward structure with the master regulator SCC regulating both
the \emph{sigf-sigg} and \emph{sigk-gere} SCCs. This second SCC also
receives signals from \emph{sigf-sigg}. The SCC \emph{lexa-rocr} receives only from the central SCC, and this functions as the input
to the circuit \emph{lexa-rocr}. Two controller nodes feed directly into the central
hub, while the third controller node connects the \emph{sigf-sigg} to
the \emph{sigk-gere} SCCs.

\paragraph{Logic circuits present in \emph{B. subtilis}'s minimal GRN}
We found only 8 logic circuits in \emph{B.~subtilis}, less than the 12 found in \emph{E.~coli}. Two of them correspond to MR circuits (toggle-switch type): the \emph{lexa-rocr} SCC and
\emph{glnr-tnra} from the central SCC. Like in \emph{E.~coli}, both of
these circuits 
present additional negative autoregulations in each gene (self-inhibitions, see SI Section~\ref{sec:SI.DetailsOnGRNs} for a discussion on the effect of these self-loops). Two other circuits are  NFBL circuits (oscillating
types): the \emph{sigk $\mapsto$ gere} SCC, which actually corresponds
to an amplified NFBL given that \emph{sigk} possess a self activation
and \emph{siga $\mapsto$ spo0a} from the central SCC, actually corresponds to a
Smolen oscillator because of its self-loops. The main difference with respect to 
\emph{E.~coli} is that we observe 4 positive autoregulation (PAR) feedback loop circuits,
possibly \emph{lock-on} circuits: the \emph{sigf-sigg} SCC and the
rest within the central SCC. All of them contain at least one gene with
an additional positive autoregulation. In \emph{E.~coli} the only
\emph{lock-on} feedback loops present were part of the
\emph{feedforward fiber} type circuits. See Table~\ref{tab:circuits}
for more details.

\paragraph{Simple directed cycles in \emph{B. subtilis}'s minimal network} The number of cycles in \emph{B.~subtilis} is a bit larger than in \emph{E.~coli}. Out of the 48 cycles, 3 correspond to the two-node SCCs and the remaining 45 cycles are located within the central SCC. As in \emph{E.~coli}, all the cycles, except for one, pass through nodes also connected by logic circuits. The one exception is the cycle formed by \emph{abrb-sigh-spo0a-abrb},
although \emph{sigh} and \emph{spo0a} do belong to logic circuits,
they are not part of the same circuits. Again, many cycles pass through multiple nodes that are connected by a logic circuit, as shown in Fig.~\ref{fig:cycles}
for \emph{E.~coli}.

\subsection{Statistical significance of the network structures observed}
\label{sec:stat sig}
 
If we find structures in graphs, we may ask whether these structures are expected in graphs of a certain type, or whether their observed numbers are unexpectedly high. In biological networks, this reflects a similar question: are the structures observed expected to appear in evolution just by chance, or are they so "unlikely" that we need to assume that they were favored by evolution for some functional benefit?  To see whether the observed structures do not only emerge by chance,
but are functionally relevant, we compared the \emph{E.~coli} and
\emph{B.~subtilis} networks to corresponding "null hypothesis"
ensembles of random graphs, following the recipe used to find
significant network motifs. {
Our random network are supposed to
represent the hypothetical outcomes of an evolution based on
mutations, but without a selection for function; edges are rewired,
while preserving some basic structure of the original networks (in
particular, the in- and out-degrees of all individual
nodes)}. Structures that appear in the real networks, but are rare
or absent in randomized networks can be assumed to be due to an
evolutionary selection, probably for specific functional advantages. To assess the statistical significance of the structures observed, we followed a standard approach: we compared our results to results from randomized networks, representing a null hypothesis. Details and some quantitative results of the analysis are given in SI section \ref{sec:SI:statisticalAnalysis}.

\begin{figure}[t!] 
    \centering
    \includegraphics[width=.95\linewidth]{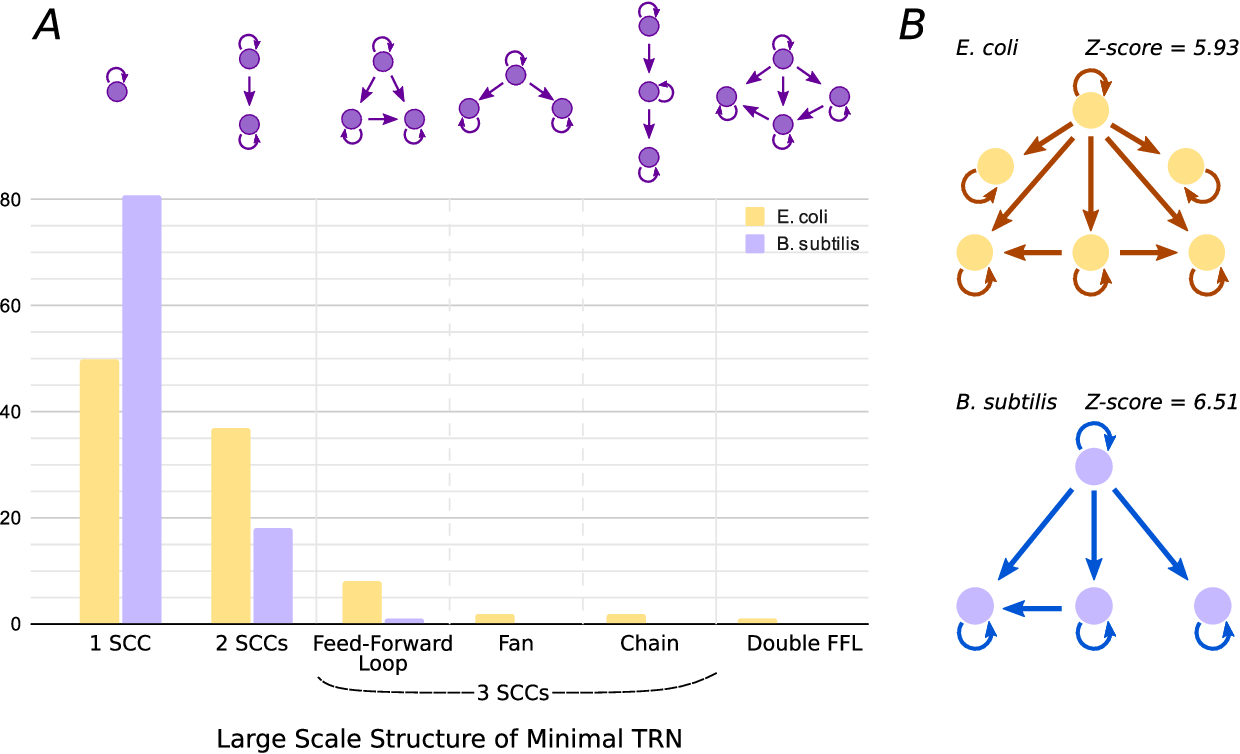}
    \caption{\textbf{Statistically significant large-scale structures in \emph{E.~coli} and \emph{B.~subtilis} core GRNs.} The structure between SCCs is
      compared to the corresponding structure in randomized networks
      with the same in- and out-degrees of all nodes (and preserving
      the edge types). \textbf{A} Histogram illustrating the
      distribution of the observed structures in the core of the
      random networks along with a sketch of the structure itself atop
      the histogram. Following the format of Fig.~\ref{fig:main}, purple
      circles represent SCCs and arrows stand for edges between SCCs. 
      \textbf{B} Structures observed in \emph{E.~coli}'s and
      \emph{B. subtilis}'s core shown with the \textit{Z-scores} of
      obtaining a structure with such number of SCCs from the
      randomized networks.}
    \label{fig:hist}
\end{figure}

The results are shown in Figure \ref{fig:hist}. Almost all the studied structures found in both bacteria are statistically significant, as shown by their Z-score on Fig. \ref{fig:hist}B (see SI Table~\ref{tab:rand} for a full breakdown). All fiber classes are significantly over-abundant in the real networks, with the exception of the simplest fiber building block of only one regulator, $|n=0,\ell=1\rangle$,
which is significantly absent. This suggests that evolved GRNs favor
more complex wiring patterns, more complex fiber building blocks than just trivial ones, that allow for richer dynamics and more
flexible control. 

Our analysis suggests that the two bacterial networks resulted from an
evolutionary selection for specific functional structures. This
concerns both the large-scale substructuring of the core into several
SCCs and the variety of small-scale circuits, which is far richer than that
to be expected with only random mutations at play.

\subsection{The bacterial minimal GRN as a computation device}

Having a clear picture of their structure, we can now describe bacterial gene regulatory networks as a computation device. 
The two primary components of computer processors are flip-flops (toggle-switches) and oscillators, both of which are also present in our minimal bacterial GRNs. Thus, the GRNs studied can be seen as computational devices in which data are stored and processed by broken symmetry circuits within the SCCs. The flip-flops in the SCCs control how fibers are turned on and off, and the symmetric fibers themselves represent clusters of genes displaying coherent synchronization in gene co-expression.  The results of these computations in the flip-flops is then transmitted to other parts of the cellular network through the fibers. The switching of the flip-flops between their stable states (i.e., zero and one) itself is controlled by a set of controller genes that regulate the SCCs (red arrows in Fig.\ref{fig:all_circuits}A for example). These controller genes can regulate one, two, or three SCCs simultaneously. 
Hence, decision making emanates from the SCCs who then turn on and off the fibers under their control. So, the SCCs are crucial as the decision-making units of the GRN.

Since all logic genetic circuits are contained in our SCCs, our reduction method preserves the entire computationally relevant circuitry of the original network. The
controlling nodes to the SCCs act as symmetry breaking nodes giving
rise to the symmetry broken circuits, more generally, as
inputs for the logic genetic circuits within the SCCs. We
would anticipate that, unlike carefully designed
and implemented circuits in synthetic biology, external inputs to
the feedback circuit structure are crucial to the computational and
biological behavior of observed circuits and to the entire \emph{computational apparatus} network and the bacteria
itself. This is why it is important to understand the topology of the whole network
and the communication between the different modules of the network
among themselves and with the extra-cellular environment.

Hence, overall, we can now describe the dynamics of the network in the following way: the computational core, i.e., the resulting minimal GRN, \emph{"executes"} a response to its input
signals coming from metabolism and from the extracellular environment; its outputs are propagated through the network in a signaling cascade-like event, through the fibration tree-like structure,  flowing outward of the computational core to the peripheries of the network.

\section{Discussion}
\label{sec:Discussion}

\paragraph{Network structures revealed by ComSym}
In gene regulatory networks, signaling is decentralized: while there are some "mighty" master regulators, there is no single central agent that ultimately controls the expression levels of all genes. However, we identified here a computational core of the network consisting of a number of "vortices" in which signals can cycle and which are connected to each other in a feedforward fashion. This core network, or minimal GRN, shows an interesting modular structure, the large-scale structure of the minimal GRN. We interpret this subnetwork as the \emph{core computational apparatus} of the network, composed of an ensemble of logic genetic circuits. Furthermore, these vortices are internally composed of various logic circuits on the smaller scale. These logic circuits perform computations that drive the dynamics, while signals then propagate through the fibers to the rest of the network. The nodes in the peripheries receive signals from the core, modify their shapes, and relay them to output nodes.


Previous works on bacterial GRNs have either focused on local motifs and larger modules arising from their integration (i.e.~FFL and Dense Overlapping Regulons) or on the co-regulation of functionally related genes. Both views ignore the arrangement of genes in the overall network structure. In contrast, our fibration analysis considers the entire network and identifies the most influential circuits on the basis of their placement at the core of the network. In the bacterial network studied, we find a rich interaction between the network's SCCs as well as the important role played by feedback loops both in determining the SCCs and the circuits within, compared to previous observations~\cite{martinez2008functional}. Following our method, all genes in the GRN can be classified with regards to the function they perform computationally in the message passing dynamics. Each gene in the GRN belongs either to a fiber or to some logic circuit or sends inputs to logic circuits. Most of the external regulators of SCCs are directed towards logic circuits serving as the inputs for their computations.

Our study into the GRNs of the model bacteria \emph{E.~coli} and \emph{B.~subtilis}, after network reduction, highlight the structures that control their dynamics, and show how signals can propagate in the network. "Computations" occur only in a subset of nodes, while the others, towards the periphery of the network, just aggregate and modulate output signals. On a large scale, the core network consists of a number of "signal vortices" in which information can cycle and which are connected by feedforward arrows. On the small scale, we found logic circuits which again may have either a feedforward shape, transmitting signals only in one direction, or contain feedbacks (allowing for permanent inner states).

Our fibers account for global signal flows in the network that reach the symmetric nodes within a fiber, which allows us to compare nodes by the global signals they receive. While the core network comprises the minimal set of nodes driving the rest of the network, the logic circuits, embedded in the core's SCCs, are located in  influential positions at the core of the network. Importantly, we not only take into account the feedback loops in these networks, a key feature that has often been missed in previous analysis, but we show them to be crucial to the decision-making and computational abilities of the network since they are what defines the logic circuits and the modules in our breakdown. 

\paragraph{Biological relevance of the structures found}
Since any network can be dissected into such structures, this raises the question whether the structures we observed are biologically meaningful or arise just by chance. A first way to test this is to check whether they are statistically significant. In fact, the numbers and sizes of vortices found in the GRNs were significantly different from the numbers and sizes expected in random graphs. 
Applying the same analysis to randomized networks with the same degree distribution as both GRNs studied, we found that these randomized networks tend to show a much simpler overall structure, consisting of only one (much bigger) vortex. In contrast to this, we found that the number of SCCs has a Z-score of 5.93  in \emph{E.~coli}, and of 6.51 in \emph{B.~subtilis} (see the SI Section \ref{sec:SI:statisticalAnalysis} for a more more in-depth discussion). Along with the number of circuits observed, also having high Z-score values (see also SI Section \ref{sec:SI:statisticalAnalysis}). The observed structuring of bacterial networks into separate vortices therefore seems to be a result of selection advantages in evolution, which suggests a biological function.

In the past, modularity and community detection, e.g.~the Louvain algorithm \cite{blondel2008fast}, have been used to partition biological networks into functionally coherent modules. In our ComSym scheme, instead, we partition the network differently.  By using a combination of fibers and SCCs we are able to break down the network to its minimal structure. The SCCs comprise structures that are functionally coherent, i.e., each SCC is mostly associated with a single type of cellular function, from sugar consumption to stress. Co-occurrence of genes in an SCC seems to be a better predictor of shared biological function than co-occurrence in a module determined by
community detection algorithms. Furthermore, SCCs control a set of fibers that -- assuming equal gene regulatory input functions -- comprise co-expressed sets of genes, which may correspond to shared biological function. 

All these (statistically significant) structures are revealed by our method and would not be visible otherwise: once the network was simplified by our graph fibration, the SCCs emerge almost by themselves, and the network structure looks suddenly simple and comprehensible! This may be a lesson for understanding other, maybe even less structured and more dynamic networks modeling other forms of collective intelligence.

\paragraph{Gene logic circuits: gene duplication and symmetry breaking}
The emergence of network motifs~\cite{milo2002network, shen2002network, mangan2003structure} has been explained by an evolutionary setting of mutation and selection, with a random rewiring of the GRN (mutation) and a selection for functional structures. While this mechanism could also explain the appearance of gene fibers, it would probably be rather slow. Another much faster genetic mechanism, gene duplication, may be able to generate fiber structures in evolution fast and almost "for free". Minor modifications in the duplicated nodes could then happen by subsequent mutations.



Gene duplication is an important evolutionary process that results in a cell having two (or more) paralogue copies of a gene (or set of genes), i.e. a set of \emph{duplicate} genes. When this process results in a set of (duplicated) genes that share the same input relations, it "creates" new fibers organically.
The duplicated set of genes belongs to a synchronous fiber. In this case, the duplication itself works as the lifting property, where the fiber node in the base is "lifted" to the nodes that belong to the fiber. In this way, gene duplication offers a plausible explanation as to why the systems studied here exhibit so many symmetries, in that so many of its nodes belong to fibers.

Not only can this explain the existence of so many nodes in fibers, but furthermore the symmetry breaking of the resulting duplicated building blocks (such as the building blocks depicted in Fig.~\ref{fig:circuitsOverview}) leads to logic gene circuits. For example, by duplicating a self-inhibiting gene in such a way that its input tree is preserved, we obtain two mutually repressive genes, which correspond exactly to the \emph{toggle-switch} flip-flop~\cite{gardner2000construction}. Further mutations to each gene that could add additional regulators, separately for each gene, would correspond to the \emph{set-reset} (\emph{S-R}) switches of the toggle switch. This is explained in more detail in the SI Section~\ref{sec:circuits hierarchy} and Fig.~\ref{fig:circuits_hierarchy}.


\paragraph{Limitations of the ComSym procedure} 
What are the limitations of our network reduction method? First, bacterial signal processing does not depend on network structure alone, but on quantitative gene regulatory input functions, with parameters that differ between genes in a fiber. Even in a simple threshold model, two genes that receive inputs from the same transcription factors would show different outputs because of different logic operations (e.g., AND versus OR) or of different activation or repression thresholds (whereby one gene may be activated faster than others). As discussed in \cite{2020predict}, our fibration analysis does not address this complexity: Instead, it assumes that all genes in a fiber share exactly the same regulation; differences in gene regulatory functions, post-translational regulation, as well as mRNA and protein degradation are not taken into account. 

How can we justify this? In the spirit of physics, a symmetrized system can be seen as a first-order approximation, revealing some important general features, in this case, of gene regulation. In this view, considering individual gene properties would be a second step in which we introduce a weak symmetry breaking or second-order approximation, which changes the predicted behavior and makes it more realistic. Although the second step is important to approach biological reality, the first step may provide important insights, while fully detailed, dynamic models of GRNs would prevent us from seeing the forest for the trees. Moreover, the successful implementation of genetic circuits in synthetic biology, capable of basic logic computations~\cite{gardner2000construction} such as memory storage and time-keeping by oscillations, suggests that analogies to logic circuits may help us understand decision making also in wild-type cells.

A second limitation of our method comes from the fact that GRNs are linked to the rest of the cell. Other forms of regulation including signaling, small-molecule regulation, and metabolic pathways also display symmetries. In theory, to trace cellular signaling flows and decision processes, a fibration analysis should include not only transcriptional regulation, but also metabolism. In our analysis, metabolites are seen as given inputs to the GRN that may modulate TF activities. In reality, metabolite concentrations depend on enzyme activities, which themselves depend on the GRN, so metabolic and regulatory networks form a large feedback loop. In a fibration analysis of the entire cell, all these networks would need to be combined. The outward pathways, which are eliminated by our \emph{'pruning'} step, may then feed back on the GRN through metabolites that can bind transcription factors as effector molecules and modulate their activities. This could be described by assuming that these metabolites can turn on and off the edges in the GRN.  Enzyme phosphorylation, which involves binding a phosphate group to activate or inactivate an enzyme, is another regulation mechanism that acts between GRN and metabolism and could be considered in a larger regulatory network. 

\paragraph{Future challenges} Despite some progress in this direction~\cite{grimbs2019system}, some challenges remain. Since metabolic reactions can have multiple substrates and products, metabolic networks have to be treated as hypergraphs \cite{battiston2020networks, klamt2009hypergraphs}. Fibrations of hypergraphs still need to be developed. Another challenge concerns the different time scales. Metabolic dynamics is much faster than gene expression dynamics, so on the time scale of gene regulation, metabolism is close to a steady state. In this quasi-steady state, metabolite levels effectively depend on enzyme activities via long-range, non-sparse interactions \cite{brgk:12}, which complicates a fibration analysis.

Finally, the ideas presented here may inspire the design of artificial GRNs with functionalities of living cells. One could start with the design of the core computational apparatus of the GRN, integrating the desired amount of logic genetic circuits, and then, based on this core network, construct "signal highways" to peripheral genes whose products perform the required biological functions. In the same vein, ComSym may also facilitate the design of minimal genomes~\cite{hutchison2016design}.


\section*{Data and code availability}

The \emph{E.coli} and \emph{B.~subtilis} networks were obtained from RegulonDB~\cite{tierrafria2022regulondb} and 
SubtiWiki~\cite{pedreira2022current}, respectively. Code for ComSym is available at: \url{https://github.com/luisalvarez96/MinimalTRN} and 
\url{https://github.com/makselab/MinimalTRNCodes}

\section*{Declaration of conflicting interests}
The author(s) declared no potential conflicts of interest with respect to the research, authorship, and/or publication of this article.

\section*{Funding}
Funding was provided by NIBIB and NIMH through the NIH BRAIN Initiative Grant \# R01 EB028157.

\section*{ORCID iD}
Hern\'an A. Makse: \url{https://orcid.org/0000-0001-6474-1324}\\
\noindent Wolfram Liebermeister: \url{https://orcid.org/0000-0002-2568-2381}\\
\noindent Luis Alvarez-Garcia: \url{https://orcid.org/0009-0008-0229-2461}

\newpage

\begin{flushleft}
{\Large
\textbf{Supporting Information}
}
\newline
\\

\end{flushleft}

\newcounter{SIsection} 
\renewcommand{\theSIsection}{SI-\arabic{SIsection}} 
\newcounter{SIsubsection}[SIsection]  
\renewcommand{\theSIsubsection}{\theSIsection.\arabic{SIsubsection}}

\newcommand{\SIsection}[1]{
  \refstepcounter{SIsection}
  \section*{\theSIsection\ #1} 
  \setcounter{SIsubsection}{0} 
}
\newcommand{\SIsubsection}[1]{%
    \refstepcounter{SIsubsection}%
    \subsection*{\theSIsubsection\ #1}%
}

\setcounter{figure}{0} 
\renewcommand{\thefigure}{SI-\arabic{figure}} 

\setcounter{table}{0} 
\renewcommand{\thetable}{SI-\arabic{table}} 


\SIsection{Specifics of gene regulation dynamics}
\label{sec:SI:GeneRegulationDynamics}

Gene expression dynamics in the GRN can be modeled using ordinary
differential equations (ODEs)~\cite{gardner2000construction,
  alon2019introduction} for mRNA and protein concentrations, assuming
known gene-regulatory functions~\cite{bbgg:05b}. 

While we will not model signal transmission dynamics, it is necessary to understand how
a network of regulatory arrows can be translated into a dynamical
system and vice-versa. This clarifies both the importance of network
structure and the importance of other, quantitative and gene-specific
details that are not represented by network structure alone. In a
simple lumped model, each gene is described by a single dimensionless
expression value, representing the protein concentration, that is, the
gene product. We consider a protein with concentration $x_i$, encoded by
gene $i$ and regulated by a set of genes $j$ expressing transcription
factors with activities $y_j$. The expression
dynamics of this model can be modeled by the following lumped ODE:

\begin{equation}
    \frac{dx_i(t)}{dt} = -\alpha x_i + F[\gamma_j f_K(y_j)] ,
    \label{eq:model}
\end{equation}

where $x_i(t)$ is the time-dependent protein concentration expressed
by gene $i$, $\alpha$ corresponds to its degradation constant
(including the often dominant effect of dilution~\cite{book-wolf}),
$\gamma_j$ is the maximum synthesis rate of the protein product and
$f_K(y_j)$ is the interaction or input function between TF $y_j$
and the binding site of gene $i$ which depends on the dissociation
constant $K$ between them.

The coupling term $f_K(y_j)$ for an individual TF $y_j$ interaction is
usually modeled as a sigmoid function such as a Hill function~\cite{gardner2000construction, alon2019introduction}, or a Heaviside
step function, its Boolean logic approximation~\cite{leifer2020circuits, oishi2014framework}. Qualitatively, the
Heaviside step function can be thought of as a Boolean logic function
in the following intuitive manner~\cite{oishi2014framework}: each
gene can take the "on" or "off" state, an activation signal being a
proxy for \emph{"turning on"} the gene while an inhibition signal is a
proxy for \emph{"turning off"} the gene.

The function $F[\cdot]$ combines the regulatory input functions of all the inputs $y_j$ of the gene $i$. They have been proposed and measured
in Refs. \cite{smsa:03,kbzd:08}. When the input function combines the
activity of several TFs, the input functions can be taken as different
logic gates. For example, a logic AND gate with $F[\cdot]$ being
a multiplicative function of its individual gene input, or OR gates
with $F[\cdot]$ being additive on the input \cite{mssz:06}.  In
general, the fact that the input function collects the activity of
more than two TFs, implies that the biological graph is actually a
hypergraph, with each input function defining a hyper-edge.  These
input functions can be more complicated forms that involve many-body
interactions where three or more inputs interact in a common way.

While these interactions are treated by hypergraphs, not by graphs, in
this paper we consider only two-body interactions captured by a graph,
since most of the results are not affected by this difference, as long
as the hyper-edge is the same for all interactions. We leave the study
of hypergraphs for a follow-up study, yet these considerations do not
affect the general conclusions about the symmetries of the GRNs.

Equation (\ref{eq:model}) contains enormous simplifications, since,
in reality, one would need to know the precise values of all
parameters that define the model to make an exact model.
These effective parameters capture everything from transcription and translation rates, protein folding, to binding and unbinding of the TF to DNA, not to mention ribosome and polymerase binding, as well as the degradation lifetimes of mRNA and proteins~\cite{book-wolf}. This set of events are lumped into one single edge representing the whole process in the GRN.
This large set of parameters are mostly unknown to the modeler trying to understand the regulatory process, therefore the
parameters used in Eq.~\ref{eq:model} are lumped parameters to account for these
phenomena effectively~\cite{gardner2000construction}.

Crucially, in order to do a precise prediction for the entire GRN
one would need to know these parameters for every edge of the network,
in our case for our \emph{E.~coli}'s GRN-operon network there are
1,835 edges, for example. It would be pretty much impossible to model
this in an exact manner using current knowledge of biology. What we do
instead is to take the most drastic approximation and take the
parameters to be the same for every edge \cite{2020predict}. This
implies that we will search for the highest possible symmetry state of
the network. Any deviation from this approximation will incur in some
network symmetry breaking by heterogeneity of parameters. Then, the
question remains whether this symmetry breaking is strong enough
to break the synchronized dynamical patterns of gene expression
observed experimentally. In fact, this approximation
can be relaxed to have the same edge parameters for the outputs of genes
involved in a fiber, but the parameters could still be different for
different genes in different fibers. {
Importantly, because the purpose of this work is to study the overall structure to determine how it impacts the information flow and decision-making process, general conclusions about the structure of the minimal GRN, the presence of logic circuits and the interactions between them still hold even if the synchronizations are reduced to only correlations and the actual dynamics stemming from Eq.~\ref{eq:model} would be more complicated.}

The uniform parameter approximation allows us to do a large
improvement in understanding the structure of the network; as will be
shown, it will reveal the ideal structure via symmetries from which
small corrections due to the heterogeneous parameter space can be
studied later on. The question that will arise then is whether the set
of equations defined on the graph may break these symmetries or
not. Our contention is that the heterogeneous parameters preserve the
structure already imposed by the graph symmetries, and this is
supported by the experimental evidence of gene coexpression patterns
(synchronization) widely obtained for these systems~\cite{2020predict}.

\SIsection{Graph fibrations and formal definitions}
\label{sec:SI:fibrations}

A (directed) \emph{graph} $G = (N, E)$ is a pair of nodes $N$ and
edges $E$, where each edge $e \in E$ is an unordered (ordered) pair of
nodes, i.e., $e = (i, j)$ for $i, j \in N$~\cite{harary1994}. It is
customary to define two functions $s, t: E \rightarrow N$ that map
each the edge to its respective source and target nodes: $s(e) = s(i,
j) = i$ and $t(e) = t(i, j) = j$. A \emph{sub}graph is a graph $g = (n, e)$ where $n \subseteq$ and $e \subseteq E$. An \emph{induced} subgraph, in turn, is the subgraph resulting from taking a subset of nodes of the original graph and \emph{all} the edges between them.


A directed graph is strongly connected if there is a path between all
pairs of vertices, in both directions. A {\it Strongly connected component} (SCC) of a
directed graph is a maximal induced subgraph that is also strongly
connected.

A \emph{graph morphism} $\varphi: G \rightarrow B$ is a mapping
between two graphs $G = (N_{G},E_{G})$ and $B=(N_{B},E_{B})$ given by
two functions mapping the nodes and edges respectively, $\varphi_{N}: N_G \rightarrow N_B$ and $\varphi_{E}: E_G \rightarrow E_B$, that satisfy 
$s_B(\varphi_E(e)) = \varphi_N(s_G(e))$ and $t_B(\varphi_E(e)) = \varphi_N(t_G(e))$
for every edge $e \in N_G$~\cite{harary1994}, a sort of commutative relation. 

Simply speaking, if two nodes are connected in $G$, they are connected in $B$ in such a way that the incidence relationship between the source node and the target node is preserved for their respective images. This means that the edges are mapped so that the edge image connects the image of the source node to the image of the
target node.

A fibration has a stronger conservation of the graph's structure than a plain morphism, since it not only preserves the incidence relation but also requires the lifting property, which preserves the input trees of all image nodes in the base.


In order to determine the fibers we start by defining the \textit{input set $I_i$} of node $i$ as the set of incoming edges $e
\in E_G$ such that $t(e) = i$ along with their respective sources $j = s(e)$. The \textit{input tree $T_i$} of node $i$ is the
input sets of the input sets taken recursively. Input trees can be finite or infinite depending on the existence of the cycles in the
network. An infinite input tree occurs anytime a node receives signals from a strongly connected component of one (an autoregulation loop) or more nodes.


A \emph{graph   isomorphism} denotes a graph morphism whose inverse is also a morphism, i.e.~a graph isomorphism is a bijective (one-to-one correspondence) graph morphism. Formally, a graph morphism $\tau: G \rightarrow C$, between graphs $G=(N_{G},E_{G})$ and $C=(N_{C}, E_{C})$, is a graph \emph{isomorphism} if and only if for every edge $(i, j) \in E_G$ there is an edge $(\tau(i), \tau(j)) \in E_C$.

Two graphs are said to be isomorphic if there exists an isomorphism between them. Fig.~\ref{fig:fibration}B demonstrates an example of two isomorphic input trees. Intuitively, the topology of these trees is exactly the same, meaning that the graphs are the same except for a relabeling of the nodes (and also the edges), and the isomorphism condition above is satisfied.

An input tree graph isomorphism defines an equivalence relationship between nodes in the graph where the equivalence classes are \emph{fibers:} nodes with isomorphic input trees. 

%

That is, the fibration collapsing nodes in a fiber into a single node in the base conserves the dynamics of the graph. Thus, in terms of an
admissible set of equations attached to the graph, fibers lead to the existence of synchronous solutions for the nodes within the fibers. This is called {\it cluster synchronization,} in this case corresponding to gene coexpression patterns~\cite{morone2020fibration, 2020predict, deville2013modular}.


We have shown before~\cite{morone2020fibration, morone2019symmetry,   leifer2020circuits, 2020predict} that the use of fibrations allows
for the breakdown of the network into its fundamental synchronized building blocks, {the \emph{fiber building blocks}. Each fiber belongs to a fiber building block, defined as the induced subgraph formed from the nodes in the fiber, and the nodes that send inputs to the fiber (the regulators). In the case where the fibers send signals back to any of its regulators (i.e. there is signal feedback from the fiber to its regulators), all nodes that belong to the shortest path from the fiber to the regulators must also be included.}

We find that these fiber building blocks can be precisely characterized by just two numbers $n$ (or $\varphi$) and $\ell$, defining the
$|n,\ell\rangle$ classes~\cite{morone2020fibration}. Here $\ell$ corresponds to the number of genes (externally) regulating the fiber, the external regulators, and $n$ the number of cycles within the fiber. $n$ can be of integer or fractal dimension (represented by $\varphi$ in these cases), depicting if the input tree's size growth can be described by an integer number or not. The latter being the case when there is a cycle between the fiber and the regulators.

The full list of fibers and their classification in the \emph{E.~coli}
GRN can be seen in the Supplementary Information File from~\cite{morone2020fibration}. For \emph{E.~coli} and \emph{B.~subtilis}, all the different fiber building block structures are classified by just 5 basic canonical structures shown in Fig.~\ref{fig:circuitsOverview}, which implies a nice reduction in complexity of these fundamental structures.

\SIsubsection{Gene coexpression and fiber building blocks.}

In GRNs, a symmetry fibration describes the synchronized expression of
genes with isomorphic input trees or, biologically, gene
co-expression~\cite{leifer2020circuits, 2020predict}. As discussed,
the underlying uniformity assumption is that the regulatory input functions of the genes, as well as their parameters, are identical between all
genes in a fiber. In biological reality, this is clearly not true:
different genes (even with the same input edges) will show different
input functions, mRNA lifetimes, etc, and so genes in a
fiber will not show strict synchronicity. However, we assume that
these genes will still show correlated dynamics and that
deviations from strict synchronicity can be described as a weak
departure from the exact synchronous state~\cite{2020predict}. Thus,
we consider a simplified picture of GRNs in which this symmetry
assumption for gene regulation functions holds.

Synchronization requires that isomorphic input trees do not experience
any significant communication delays which would cause asyncronicity
and that the constants involved in Eq.~(\ref{eq:model}) are
approximately the same for all the genes in a fiber, as
discussed. This is not difficult to satisfy in GRNs where interactions
depend mainly on the TFs and not so much on the binding site of the
regulated genes. Small variations resulting from mismatching
parameters appear to result in weak symmetry breaking, creating a
slight reduction in the synchronization and correlation of expression
levels within the fibers~\cite{2020predict}.

The co-expression patterns obtained here are more general than
traditional co-expression patterns in operons and regulons. Indeed,
all genes in an operon with a single common promoter are in the same
fiber. Such operons can be thought of as examples of \emph{"trivial"}
fibers, see Fig.~\ref{fig:circuitsOverview}. The same applies to
nodes that belong to only one regulon and share the same regulatory
TF, they form a \emph{"trivial"} fiber since they are only regulated
by one gene in an identical manner. Generally speaking, symmetry
fibrations allow us to find not only these trivial fibers but also
broader co-expression patterns among genes that are far away in the
genome and more complex patterns of synchronization. Particularly
interesting fibers, shown in Fig.~\ref{fig:circuitsOverview}, are the
Fibonacci fibers where the presence of a feedback loop between the
fiber and its regulator (forming a SCC), plus a self-loop in the fiber
produce an input tree with a branching ratio of fractal dimension, as
seen in Fig.~\ref{fig:circuitsOverview}. Other complex fibers are
composite multi-layered fibers, in which nodes that are not regulated
by the same node are still synchronized because their regulators
belong to a fiber.

It is important to note that the theory of fibrations only predicts
the existence of these symmetric synchronized solutions.  But not all
synchronous solutions must be symmetric. More importantly, symmetries
do not guarantee that the synchronous solution will be stable, and
solutions may (in theory) be dynamically unstable~\cite{golubitsky2006nonlinear, stewart2003symmetry}.

Thus, fibrations do not cover all dynamics in the original network,
and guarantee existence but not stability.  Fibrations cannot
guarantee that these symmetric solutions are actually relevant dynamic
attractors and not unstable. The stability of the synchronous solution
needs to be studied a posteriori, and it depends on the particular
type of ODE used to describe the dynamics.  Thus, different
stabilities can be obtained for different models, whether we use a
Hill function or a linear interaction term or a step function, for
instance, in the ODE. In fact, there are very interesting bifurcations
that can exist for a given dynamical model, and bifurcation can be
symmetry preserving or symmetry breaking.  Each fiber needs to be
investigated separately for each model. The stability and bifurcation
analysis of the circuits found in bacterial GRNs are
investigated in detail in Ref.~\cite{stewart2023}.

In summary, by studying the input trees of all nodes in a graph we can
determine all the symmetries of the network in terms of signal
processing. Specifically, the symmetries of a network are given by the
equivalence relations induced by input tree isomorphisms. Therefore,
symmetric paths can be removed while preserving the flow of signals,
which is concretely related to the concept of fibration~\cite{boldi2002fibrations, deville2013modular} and its implications
for dynamics in biological networks~\cite{morone2020fibration,
  leifer2020circuits}.

\SIsection{The ComSym analysis step by step}
\label{sec:SI:ComSymStepByStep}
 
\SIsubsection{Step I: The symmetry fibration -- Collapsing a graph into its minimal base by surjective fibrations}

The main reduction process is the application of surjective
minimal fibrations, or {\it symmetry fibrations}, which reduces the
original network to its minimal base. This is the fibration that
collects the maximal symmetry. We first identify all isomorphic input trees (see SI and Refs.~\cite{morone2020fibration, monteiroAlgorithm} for a discussion on algorithms), obtaining the least amount of fibers, or colors, hence also receiving the name minima balanced coloring. Afterwards, all fibers can be collapsed into a single representative node, obtaining as a result the minimal base. This base network represents the reduced effective model network for the original system and has the same original signaling flow but with no redundant pathways. Since nodes in the same fiber share identical inputs, these are not changed; however, they generally have different outputs. When the collapsing is done, all previous outputs from all nodes in the fiber must be "rewired" so that their new source node is now the collapsed fiber-node. 


DeVille and Lerman~\cite{deville2013modular} have shown that any surjective fibration
$\varphi: G \rightarrow B$ induces synchronization on nodes $i$ and $j
\in N_{G}$ if $\varphi(i) = \varphi(j) \in N_{B}$. Thus, for the case
of a symmetry fibration, all nodes within the same fiber are
synchronous.  This guarantees that the dynamics for both the original
network and the base network are the same, all the collapsed nodes
dynamics are identical to the representative node they were collapsed
into.

Importantly, this reduction is valid for any signal-processing network when a substantial reduction in network size is desired without losing signaling flow~\cite{morone2020fibration}.

\SIsubsection{Step II: Injective Fibrations} 
\label{sec:inj}

An injective fibration can help us formalize the intuitive notion that
under certain conditions the dynamics of an entire system may be
driven by only a subset of its constitutive elements. This is
formalized by \emph{Lemma 5.2.1} in DeVille and Lerman,
Ref.~\cite{deville2013modular}, where it is shown that for an
injective (one-to-one) fibration $\varphi: G \rightarrow G_2$ the
dynamics of the bigger graph $G_2$ is driven by the dynamics of the
smaller graph $G$. In fact $G$ is a subgraph of $G_2$. This can be
seen on Fig.~\ref{fig:fibration} with the injective fibration from $G$
to $G_2$.

The emphasis here is on the injective nature of the map $\varphi$. For
a mapping to a larger graph to be a fibration, like the one shown in
Fig.~\ref{fig:fibration}, it must satisfy the lifting property. This
requires that all edges in $G_2$ whose target node is an image
from a node in $G$ (i.e. $1',2',3',4'$), can be uniquely \emph{lifted}
to an edge in $G$. This implies that no new edges are allowed to
target any of the nodes of the original graph (that is, $1',2',3',4'$),
satisfying the lifting property.
As a consequence, all the added nodes in $G_2$ (i.e. $5', 6', 7'$)
must strictly be only targets of the image nodes. Hence, signals flow
only outward from $G$, and therefore the dynamical state of the outer nodes
is driven by the dynamics of the original smaller graph G. In other
words, the subgraph $G$ of $G_2$ drives or controls the entire
dynamics of $G_2$.  This, in turn, guarantees that the dynamics of the
original graph is preserved in the image graph $G_2$.

However, the issue for us is how to
reverse this process to obtain the driver subgraph $G$ given
$G_2$. We propose that the $k$-core decomposition of a
graph is the tool that allows us to perform the \emph{inverse}
injective fibration to find the driver subgraph of the network $G$. This will allow us to distinguish between the nodes that shape the network dynamics and those that are just being driven.

\paragraph{The $k$-core decomposition reveals the key nodes that drive the network's dynamics}
\label{sec:k-coreDecom}

\begin{figure*}[!ht]
    \centering
    \includegraphics[width=.75\textwidth]{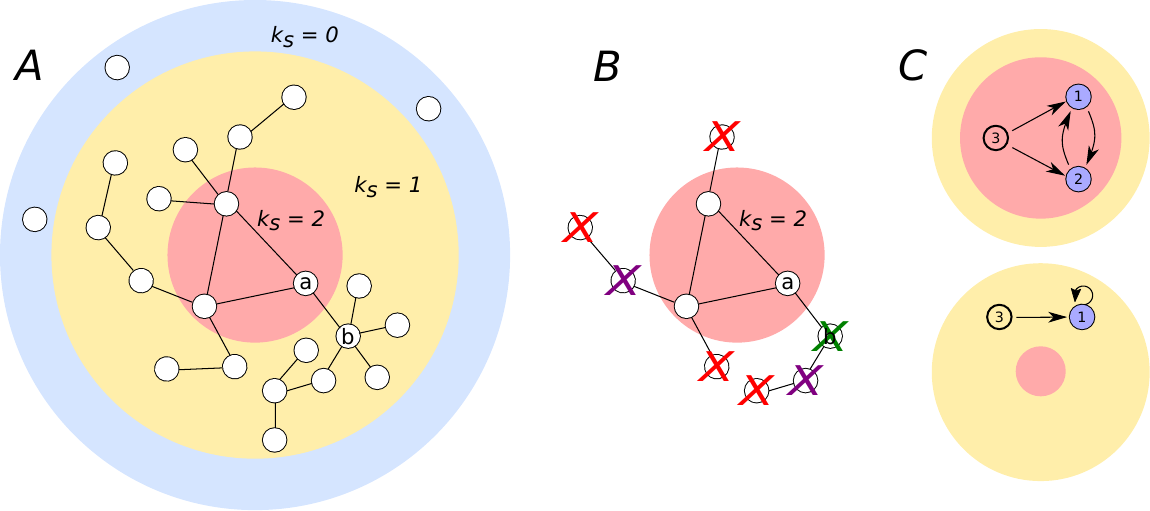}
    \caption{{\bf $k$-core decomposition}. 
    \textbf{A} Schematic drawing of the $k$-core decomposition of an undirected
      network. Even though node $b$ has a higher degree than node $a$,
      it is connected to nodes with smaller degree and has therefore a
      smaller coreness than node $a$. 
      \textbf{B} Example of how to
      obtain the $k = 2$ core of the network in \textbf{A}. First, the
      nodes with degree less than $2$ are removed. The remaining nodes
      are the ones shown, successive iterations remove the nodes
      crossed with colored \emph{x}. Different colors stand for
      successive iterations: red is the first iteration, the second
      one is purple and the last one green.  
      \textbf{C} The bottom
      network corresponds to the symmetry fibration of the network on
      top. The SCC of nodes $1$ and $2$ is also a fiber, after
      collapsing the fiber, the SCC is lost, becoming a single node with a self-loop. The collapsed
      network then becomes an acyclic graph, without a $k_{\rm out} = 1$ core.}
    \label{fig:kcore_SI}
\end{figure*}

Formally, the $k$-core of a network is the maximal induced subgraph,
where the degree of each node within the $k$-core is at least $k$.
It consists of "peeling off"
layers, the $k$-shells of a network, by assigning the coreness index
$k_{s}$ to each node, corresponding to the respective shell they
belongs to~\cite{malliaros2020core}. The $k$-shell corresponds to the
set of nodes with coreness $k_{s} = k$.  The coreness of a node is
given by $k$ if it belongs to the $k$-shell, that is, the $k$-core but
not to the $(k+1)$-core. The lower the coreness $k_{s}$ of the node,
the more peripheral it is~\cite{kitsak2010identification}, as can be
seen in Fig.~\ref{fig:kcore_SI}. 

For example, since nodes having a degree of at least $0$ belong to the $k =
0$ core (all nodes), node $b$ belongs to the $k = 0$ core
and to the $k = 1$ core, as shown in Fig.~\ref{fig:kcore_SI}A, but
does not belong to the $k = 2$ core, so its coreness is $k_s =
1$.

Taking the $k$-core of a network corresponds to removing all nodes
with degree less than $k$, calculate the new degrees in the new
subgraph, and remove again nodes with degree less than $k$,
iteratively until all remaining nodes have a degree of at least
$k$. A node with coreness $k$ has a degree of at least $k$,
connecting to other nodes within the same core. 

The coreness of a node captures the degree of the nodes to which it is connected. For example, node $b$ in
Fig.~\ref{fig:kcore_SI}A has a smaller coreness ($k_s = 1$) than
node $a$ ($k_s = 2$), although $b$ has a higher degree of $5$
compared to the degree of $a$' of $3$. This is because $b$ is mostly connected to nodes with a degree of $1$, in contrast to $a$ which connects
to nodes with a higher degree, making it more
\emph{"influential"}~\cite{kitsak2010identification} or central.

To apply this concept to bacterial GRNs we must first extend it to directed
graphs~\cite{malliaros2020core}, where each node now is characterized
by an \emph{in-degree} and an \emph{out-degree}, instead of the usual
simple degree in undirected graphs.  However, since we are interested
in the direction of outward signaling flow, given that this case
corresponds to the direction of regulation from one gene to another,
we only need to consider the out-degree of each node. We are now
interested in the $k_{\rm out}$-core, which corresponds to the maximal
induced subgraph where every node has at least an \emph{out-degree} of
$k$. This means that in this case we are iteratively removing
nodes with less \emph{out-degree} than $k$. See for example in Fig.~\ref{fig:fibration}D the $k_{out}$ core of network $G$. 

In our case, we are only interested in genes that send any signals to
other genes, and we want to remove the ones that do not. Hence, we want to
remove the $k_{\rm out} = 0$ shell of the network and obtain the
$k_{\rm out} = 1$ core. 
In a sense, we are trimming the frayed ends, or loose ends, of the
network to unravel its core, this core network is the minimal GRN. 


When collapsing only to the $k_{\rm out} = 1$ core without applying the
symmetry fibration, one also finds a subset of nodes that drive the entire
graph. However, this may not necessarily yield the \emph{minimal}
subset, as there may be some redundancy in the signaling
flow of the resulting network.
In the case of a 2-node
SCC whose nodes also belong to the same fiber, shown in
Fig.~\ref{fig:kcore_SI}C, given that both nodes $1$ and $2$ have an
\emph{out-degree} of one, they belong to the $k_{\rm out} = 1$ core of the
original network. However, when applying the symmetry fibration first,
they collapse into a single node with a self-loop and no
\emph{out-degree}, which implies that they no longer are part of the
$k_{\rm out} = 1$ core.

Hence, as input tree isomorphisms give us the correct way to collapse a network while preserving signaling flow, the $k$-core decomposition, and in particular the $k_{\rm out} = 1$ core,
gives us the correct \emph{inverse} fibration to reduce the network to its core network, its minimal set of nodes driving the dynamics.

\SIsubsection{Step III: Strongly Connected Components and their interactions}

In undirected networks, the $k = 1$ core contains the giant
connected components of the network, this can be seen in
Fig.~\ref{fig:kcore_SI}A. Analogously for directed networks, the
$k_{\rm out} = 1$ core is composed of the \emph{strongly} connected
components, the nodes that feed them signals and the connections between them. As shown in
Fig.~\ref{fig:fibration}D the $k_{\rm out} = 1$ core contains the SCC
involving nodes $1$ and $2$ as well as node $4$ that sends signals to
it. As a corollary, all acyclic directed graphs have a null
$k_{\rm out} = 1$ core.


In this step, what we wish to find is how to break the minimal network into its components, i.e., how the SCCs interact with each other and what structure of the minimal network emerges from this. This will give us what we call the \emph{large-scale structure of the minimal network}. After this, we "zoom in" onto the SCCs to analyze the \emph{small-scale structures} inside of the SCCs in Steps IV and V.

\SIsubsection{Step IV: Broken symmetry circuits: hierarchy and identification}
\label{sec:circuits hierarchy}

We find that circuits with broken fibration symmetry act as circuits
performing basic logic computations in the GRN~\cite{leifer2020circuits}. These computations are of two types: memory storage and timing via oscillations.

\begin{figure}[!ht]
    \centering
    \includegraphics[width=.75\linewidth]{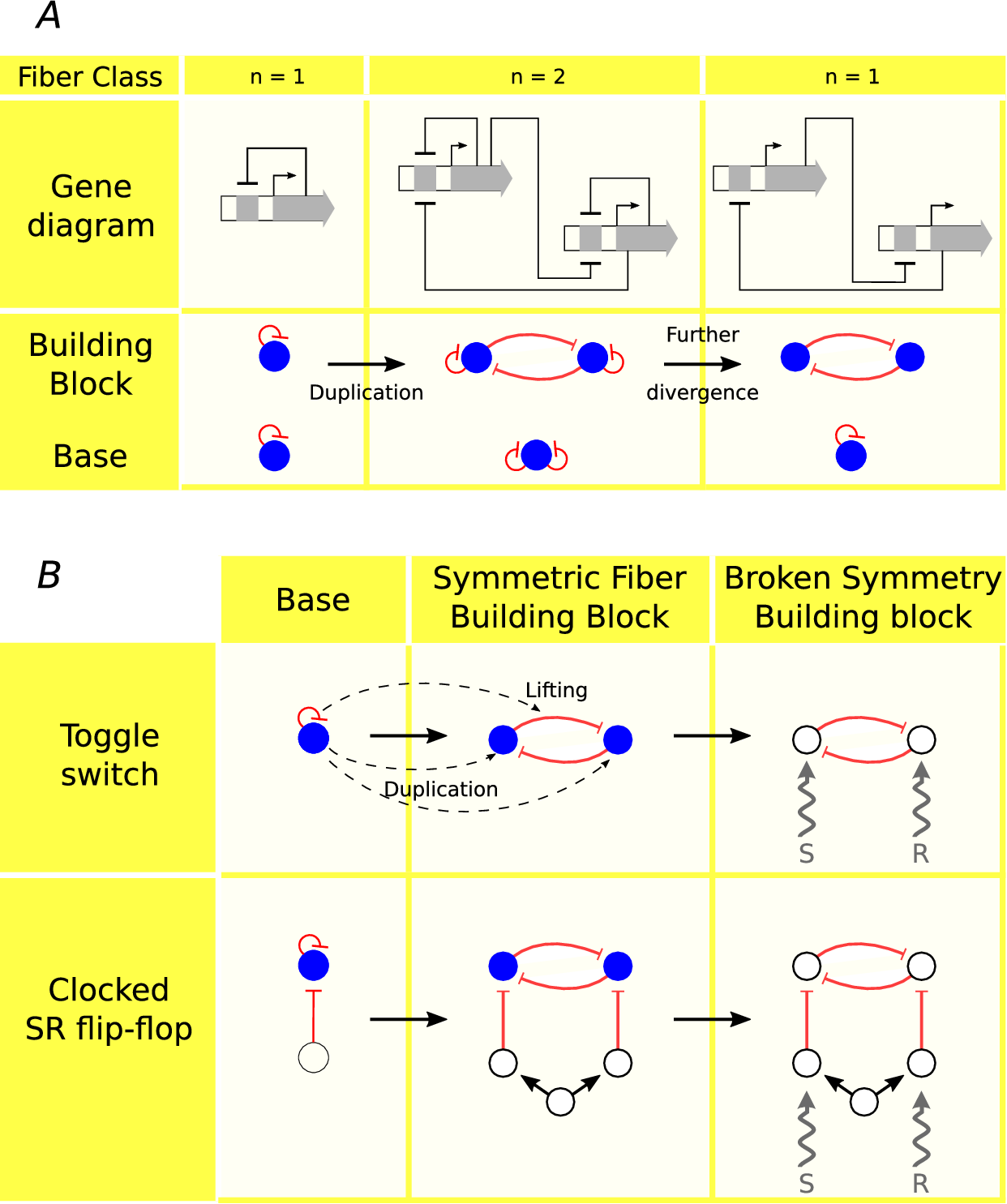}
    \caption{
    {\bf Broken symmetry circuits and duplication process.} Starting with an initial simple structure a gene duplication process can create a symmetric fiber structure, analogous to a lifting process. \textbf{A} The structure resulting from the duplication process may need to undergo further modifications. \textbf{B} The circuit formed depends on the initial structure that is replicated, shown at the bottom.  For both circuits, the
      symmetric form is shown on the center, while the broken symmetry
      process is shown on the right. The \emph{set-reset (S-R)} inputs
      break the symmetry by sending different regulations to an
      otherwise synchronous pair of nodes. The \emph{S-R} inputs can
      be any different nodes sending different signals or can be a same node sending different signals to the nodes in the fiber.}
    \label{fig:circuits_hierarchy}
\end{figure}

\paragraph{Gene duplication could explain the rich symmetries in GRNs}


Gene duplication is an important and major process in the evolution of a genome in which entire pieces of the chromosome are duplicated, resulting in the cell having two paralogue copies of a set of genes. This duplicates not only the gene but also its promoter region, the DNA sequence adjacent to the gene, where the binding sites to which the TF can attach are located.

In this way, gene duplication could work as a driver for fibration symmetries in transcription regulatory networks. By duplicating a gene in such a way that the paralogue genes share the same input relations, creates a fiber since the duplicated genes would have an isomorphic input tree. In this instance gene duplication works exactly as a lifting property, starting from the initial network and going into a new one with more symmetries and bigger fibers. Further mutations in the coding region then cause the two original copies of the gene to perform different functions, and thus diversify the bacteria's functions while creating bigger fibers. 

Logic circuits are identified by starting with a {symmetric circuit, which originates from a fiber building block,} and breaking
its symmetry by adding extra edges acting as regulators that break the symmetry in the fiber structure.  
Breaking of symmetry can be the result of a process of gene duplication that
starts with the symmetric base of a fiber building block, which is duplicated to a (still) symmetric structure. This new structure may then incorporate new regulations that result in the symmetry breaking on this duplicated circuit, as seen in Fig.~\ref{fig:circuits_hierarchy}B. 
Crucially, if the result of a duplication process respects the lifting property and preserves the fibration symmetry (see also Fig.~\ref{fig:circuits_hierarchy}A), this new obtained symmetric structure is prone to obtain new regulations that will then result in symmetry breaking. 

{Strictly speaking, the duplication process of a base does not necessarily produce a symmetric structure or a structure with the same fiber classification (see Fig.~\ref{fig:circuits_hierarchy}A). For example, if a self-inhibiting gene is duplicated into two genes, a precise duplication of the self-inhibition should produce a structure in which the two paralogue genes mutually repress each other as well as also self-inhibit themselves, see Fig~\ref{fig:circuits_hierarchy}A. The original structure corresponds to a $n = 1$ FFF, while the new duplicated structure would in fact correspond to a $n = 2$ instead, which accounts for the two interactions that the new paralogue genes receive. However, the initially duplicated structure may undergo further divergent modifications, like the removal of the self-inhibitions in this case, that result in a structure that actually preserves the same fiber classification since the input relations are the same as in the original structure. Hence resulting in a structure with the same original dynamics that respects the lifting process. This new structure, when collapsed via a symmetry fibation gives as a result the actual original structure that got replicated. This further process of modifications after duplication can be thought of as a divergence process of the duplicated structure. In this case, there are now multiple synchronous genes where before there was only one.}

The duplication process plus divergence \emph{'opens up'} the base of a fiber building block, {resulting in, at least, a pair of synchronous genes. In this instance, this process is} identical to
how the lifting property \emph{'opens up'} a fiber from the base into the full set of synchronous genes. {The identification of the liftying property as a duplication process could actually account for a possible explanation of the emergence of fibration symmetries in these GRNs and why biological networks present such a rich symmetric structure (see for example Fig.~4 in~\cite{morone2020fibration}.)}

\paragraph{Symmetry breaking gives rise to gene-logic circuits}




Duplicating a self-inhibiting gene in such a way that the lifting
property is respected implies that the duplicated circuit results in
two mutually repressed (MR) genes forming a two-node negatively
auto-regulated (NAR) fiber (a fiber that auto-regulates itself in an
inhibitory manner) as the one shown in Fig.~\ref{fig:circuits_hierarchy}A. This duplicated circuit is still a
symmetric circuit {whose dynamics are} not the same as the negative
auto-regulated loop in the base, since each gene receives two inputs
in the duplicated circuit, and the base receives only one. So, this
duplication event does not conserve lifting, and the duplicating genes
cannot be fiber to the base.
If the genes further "lose" the
autoregulations, then the resulting circuit is the lifting of the
base, and the fibration can be applied back and the dynamics are conserved.

When {different regulators are added to each gene}, then the symmetry
is broken, and the resulting circuit is a bistable switch known as a
toggle-switch. This circuit corresponds to a structure analogous to a
\emph{flip-flop} in electronics~\cite{tanenbaum2016structured} with
the different inputs for each gene being the \emph{'set' (S)} and
\emph{'reset' (R)} switches, see
Fig.~\ref{fig:circuits_hierarchy}B. This circuit is the {bistable}
toggle-switch~\cite{gardner2000construction, leifer2020circuits} that
stores one bit of information given that {as a bistable switch} it has
two possible stable and reciprocal states.  {One of these stable
  states correspond to} one gene being expressed and the other one
inhibited, and its reciprocal case {being the other stable
  state}. Even if the symmetry is restored and both inputs return back
to being identical, {the state of circuit remains unchanged, hence
  storing the previous state as one bit of information.} It is
possible to switch between the two states by toggling the \emph{S-R}
inputs to the circuit, hence why this is referred to as a {bistable}
two-way switch.

{The most simple form of a bistable switch correspond to this flip-flop, the two genes inhibiting each other as explained before (Fig.~\ref{fig:circuits_hierarchy}). If other forms of self-regulation are added to this basic structure the resulting structure still posses two different stable states, and such it still remains as bistable switch, albeit the solution of their dynamics are altered from the initial~\cite{stewart2023}.}

An analogous symmetry breaking process with the circuit generated from
a Feed-Forward Fiber (FFF) building block (from Fig.~\ref{fig:circuitsOverview}) instead, shown on the bottom
part of Fig.~\ref{fig:circuits_hierarchy}B, gives a circuit
resembling a clocked \emph{SR flip-flop}~\cite{leifer2020circuits,
  tanenbaum2016structured}, which basically acts as a more complex
\emph{flip-flop} or toggle-switch. A FFF consists of a Feed-Forward
Loop (FFL) motif (a three gene structure of genes X, Y and Z, where
gene X regulates both Y and Z, and gene Y also regulating
Z)~\cite{mangan2003structure} with an extra self-regulation on gene
Y. This self-regulation in Y induces synchronization on both genes Y
and Z, since it induces their input trees to be
identical~\cite{morone2020fibration, leifer2020circuits}. Depending on
the signs of both regulations by X and Y, the inputs received by gene
Z can be coherent or incoherent, classifying the FFLs as either
coherent and incoherent~\cite{mangan2003structure, alon2019introduction}. 
Analogously, the FFFs are classified as SAT-FFF (satisfied) or UNSAT-FFF (unsatisfied)~\cite{leifer2020circuits}, respectively. UNSAT-FFF meaning that the two inputs are incoherent between each other and thus may produce an oscillating dynamic for the synchronized genes from their dynamical equations, given the competing and contradictory inputs~\cite{leifer2020circuits}, meaning this circuit works as a clock component.

This process can be continued with more complex fiber building block
structures (from Fig.~\ref{fig:circuitsOverview}), leading to a hierarchy
of broken symmetry circuits as shown in
Ref.~\cite{leifer2020circuits}, for example with Fibonacci fibers to
give Fibonacci circuits, analogous to a JK flip-flop in
electronics. However, these more complex circuits were not observed on
the bacterial GRN studied in this work, but on more complex species,
like yeast and human.

\SIsubsection{Step V: Identifying simple directed cycles}

{As discussed earlier, the SCCs can be thought of as "signal vortices" where the signals can cycle. These vortices are relevant because they are constituted of feedback loops (a form of cycle), without which the GRN's computational capacity would be drastically reduced to a "combinatoric" only nature. Indeed, feedback loops are what allow for the more complex dynamics of the logic circuits discussed above and why they are embedded in the SCCs. The SCCs, being these "signal vortices", in fact correspond to a very intricate cobweb of feedback loops. }

After reducing a network via the $k$-core decomposition as described in
section~{\ref{sec:inj}}, all the remaining nodes have
an output degree of at least one, i.e. all "null pathways" (pahtways ending at nodes
with no-outputs) have been removed. This means that the SCCs are left intact
during the $k$-core decomposition. This is important not only given that all the logic circuits belong to the SCCs but due to the interconnectedness of the nodes within a SCC, different circuits on a same SCC will therefore be intertwined with each other in non-trivial ways. In a sense, the SCCs consist of a very complicated entanglement of logic circuits.

In order to study the interconnectedness and interplay between these circuits, as well as to understand the structure of the SCCs
we take a look at the independent simple cycles present
in the minimal GRN. 
Such cycles are important not only because they connect different logic circuits and give structure to the SCCs but because, in itself, a cycle represents a form of longer-term memory, as a messaging signal is looping around the cycle.

For this, we look for all the independent simple directed cycles in the network, where a
simple directed cycle is a closed path that crosses each node just
once, except for the initial/final node. We search for a list of independent simple cycles, the cycle base, as all other cycles can be constructed by a sum of these, given that they span the cycle vector space~\cite{kavitha2009cycle, gruber2012digraph}.
This is related to the
concept of Betti Numbers in simplicial
homology~\cite{munkres2018elements}, where the $k$-th Betti number
illustrates the number of $k$-dimensional holes in a space. In this sense, a cycle describes a type of topological hole. In particular, in the
case of undirected graphs, the first Betti number $b_0 = |CC|$ where $|CC|$
corresponds to the number of connected components, while the next
Betti number $b_1 = |E| - |N| - |CC|$ gives the number of independent simple
cycles~\cite{berge2001theory}.

However, in order to construct this cycle base, cycles must be allowed to traverse a directed edge in the opposite direction (by having a negative contribution). For example, in a feed-forward loop motif it is possible to construct a cycle if one traverses at least one of the edges in the opposite, "negative", direction~\cite{kavitha2009cycle}. However, this does not make biological sense, and so we work with an "incomplete" base of cycles only allowed by existing pathways in the network, whose combinations span only the biologically existing pathways in the network. Under this mathematical restriction, in
directed graphs, determining the number of cycles is a
$NP$-complete problem~\cite{gruber2012digraph}. See SI for more details on the algorithm for finding the cycles.

Since a self-loop or buckle just connects a node to itself in a cycle of length 1, they don't really contribute to the computational machinery and hence, as throughout, we ignore them for our cycles analysis.

\SIsection{Details of the Gene Regulatory Network of \emph{E.~coli}}
\label{sec:SI.DetailsOnGRNs}

\begin{figure*}[!t] 
    \centering
    \includegraphics[width=.65\linewidth]{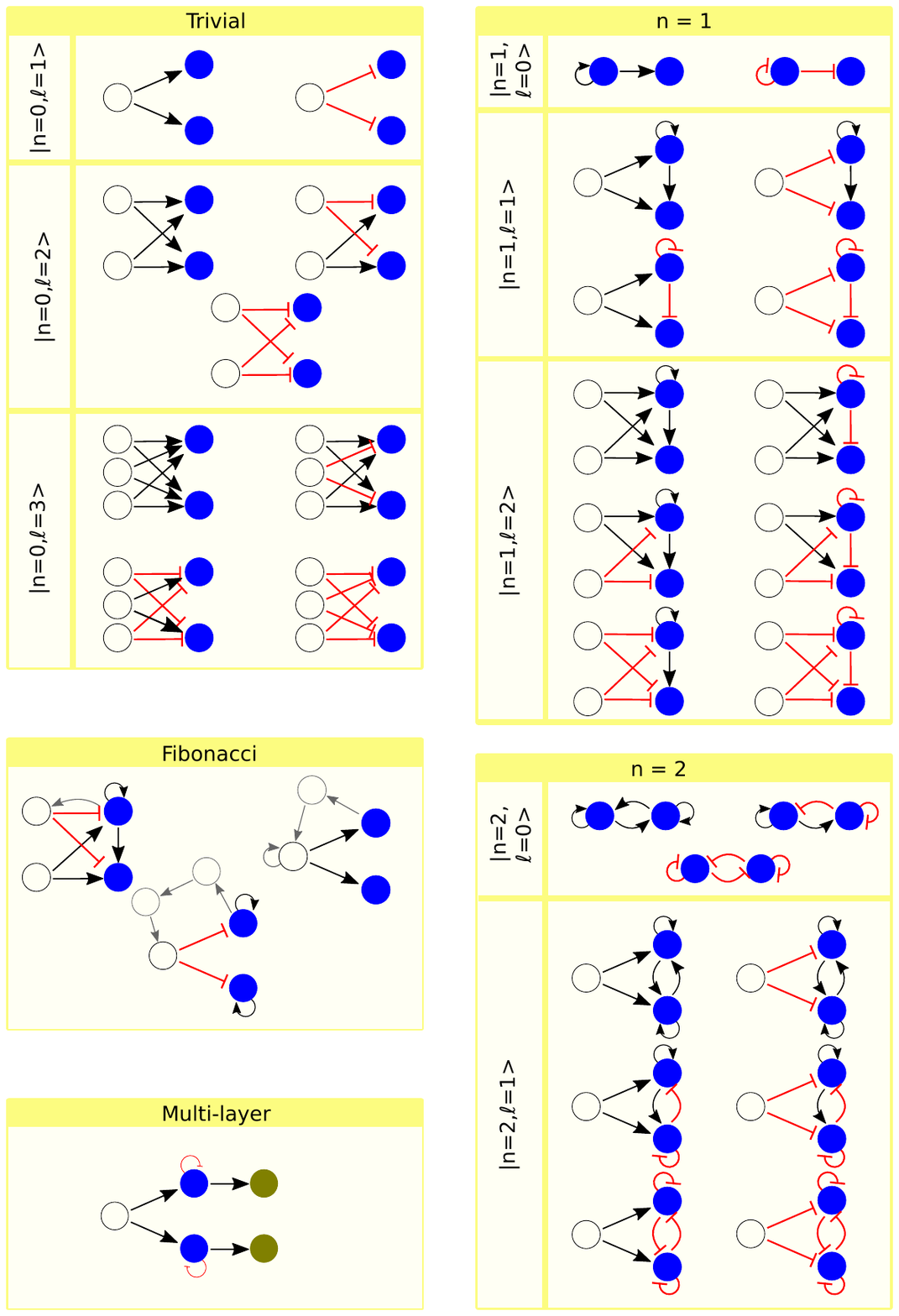}
    \caption{{\bf Possible circuits of the observed building block
      structures from \emph{E.~coli} and \emph{B. subtilis}}. We take
      the observed $|n, \ell\rangle$ classes in both bacteria and show
      all the possible combinations of activation and inhibition
      regulations for each class, i.e. all possible configurations
      which would still have a symmetric pair of nodes. In order for
      symmetry to occur the pair of synchronous nodes must receive an
      identical input tree. For the $n = 1$ cases, this can occur by a
      node with a self loop also regulating the other node in the
      fiber as shown, or alternatively by both nodes having a
      self-loop.}
    \label{fig:poss fibs}
\end{figure*}

\begin{figure*}[!ht] 
    \centering
    \includegraphics[width=.7\linewidth]{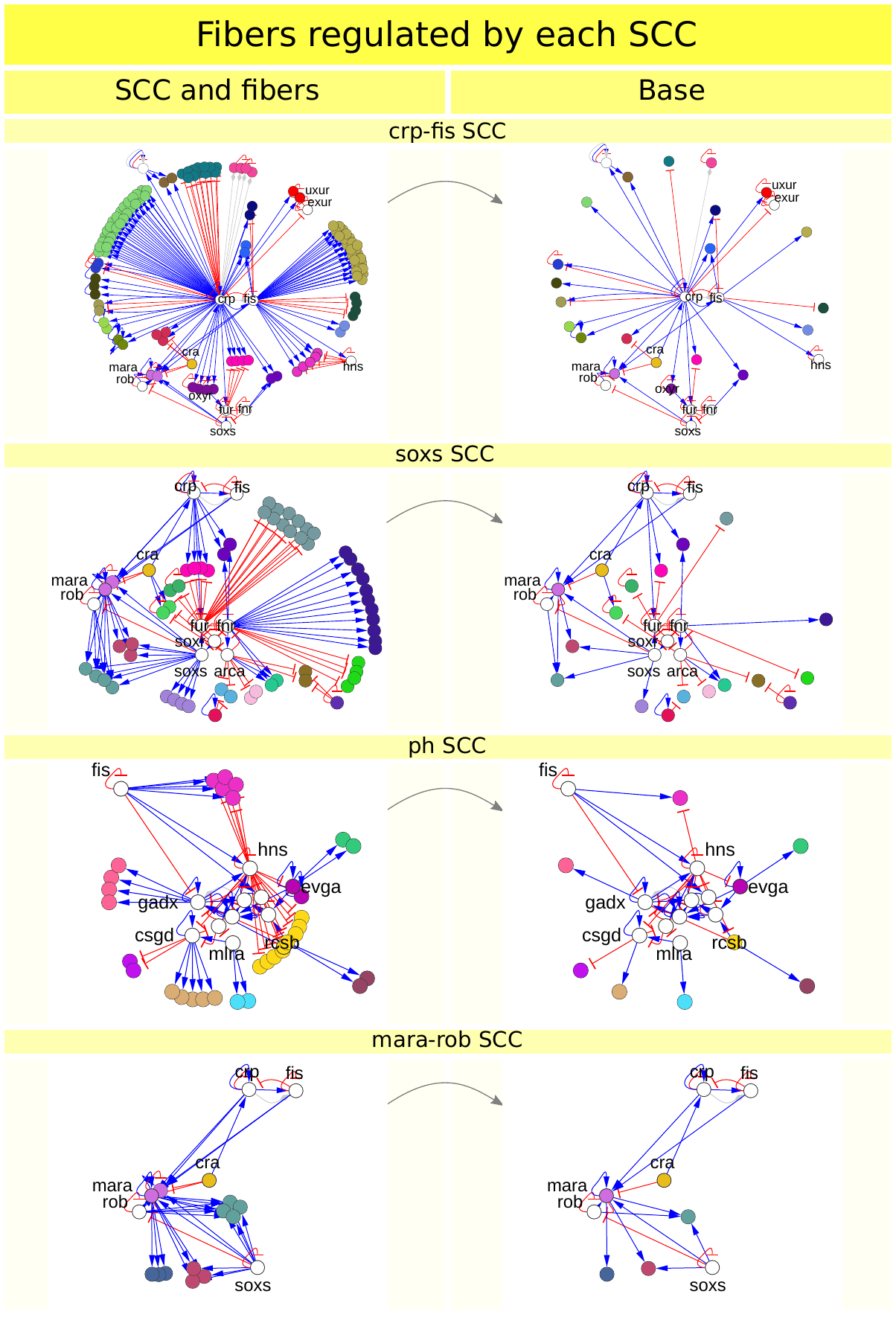}
    \caption{{\bf Example of a symmetry fibration applied to the SCCs and
      the fibers they regulate in \emph{E.~coli}'s operon GRN.} On the
      left, genes that belong to the same fiber share the same
      color. On the right all the fibers have been collapsed to a
      single representative node per fiber. Nodes shown with labels
      are nodes belonging SCCs or regulators to them. It is
      interesting to note that quite a lot these fibers are actually
      regulated by not just the nodes of one SCC but by two SCCs. For
      example, in the second row for the \emph{soxS} SCC there can be
      seen two fibers between this SCC and the \emph{mara-rob} SCC in
      greenish and brownish colors respectively. The same thing occurs
      with two other fibers shared with the \emph{crp-fis} and
      \emph{soxS} SCCs in pinkish and purple colors. Out of the 6 SCC,
      the \emph{galR-galS} and \emph{uxur-exur} SCCs do not regulate
      any fibers of their own and thus do not appear on this figure.
      Blue edges with arrows represent activation, red edges with bars
      represent inhibition and grey edges with rhombuses represent
      dual regulation }
    \label{fig:surjmin}
\end{figure*}

\paragraph{Details on signal vortices}

In Fig.~\ref{fig:surjmin} we can see the fibers regulated by all the SCCs
and their corresponding bases. Perhaps unsurprisingly, we see that
\emph{crp-fis} regulates the biggest fibers.  Also of interest is that
there are some fibers that are jointly regulated by two SCCs and that \emph{uxuR-exuR} and \emph{galR-galS} do not regulate any fiber
of their own although they do regulate other genes and operons, just not synchronously. 

The importance of the carbon \emph{crp-fis} SCC is exemplified in the
minimal network shown as the result of the reduction process. 
When considered as an
effective network, where each SCC is a super-node, the minimal GRN is
a tree structure between the SCCs, i.e. without cycles and feed-forward.  At the root
of the tree is the carbon SCC which works as a type of master
regulator, controlling the rest of the SCCs. All of these SCCs are
regulated by different genes. For example, \emph{cra}-fiber and
\emph{ihfAB} regulate the carbon SCC.
The \emph{galR-galS} and the \emph{uxuR-exuR} SCCs receive signals
from \emph{crp-fis}, but do not receive or send signals to the rest of
the SCCs. They compute their state solely based on the input of \emph{crp-fis} SCC and then send their corresponding outputs to the
genes that they regulate.

The other SCCs are arranged in the shape of a feed-forward motif:
\emph{crp-fis} SCC feeding the \emph{soxS} SCC and \emph{pH} SCC,
the \emph{pH} SCC also receives regulation from the \emph{soxS} SCC (carbon~$\rightarrow$~ph; carbon~$\rightarrow$~stress; stress~$\rightarrow$~ph) and
another similar structure, with \emph{marA-rob} SCC instead of \emph{pH} SCC: being regulated from both \emph{crp-fis} and
\emph{soxS}. The general structure of the computational minimal core
of the \emph{E.~coli's} GRN corresponds to two \emph{feed-forward}
structures between the SCCs.

\paragraph{Details on gene circuits}

We found three circuits resembling toggle switches: each of them
consists of two mutually repressed (MR) genes but with different
self-regulations. Among them are the SCCs \emph{galR-galS} and
\emph{uxuR-exuR}. For both of them, their only input gene is
\emph{crp}, which means that it could function as the logic \emph{S-R}
input selecting the state for the circuits. Both of these circuits
present additional negative autoregulations in each gene (see
Fig.~5 in main text), so their dynamics requires further study, as compared to the
classic toggle switch design.

In the case of \emph{uxuR-exuR} cycle, with the regulations of \emph{crp} this circuit then becomes the \emph{Mutual Repression Network with Negative Autoregulation} studied by Hasan {\it et al.}~\cite{hasan2019improvement}. This circuit can show two distinct stable states and may therefore serve as a memory: when \emph{crp} is active, it can induce a state in which \emph{uxuR} is expressed while \emph{exuR} is repressed. The third possible toggle-switch-like circuit is between \emph{csgD-fliz} in the \emph{ph} SCC, shown in Fig.~\ref{fig:all_circuits}, with numerous possible ways for its symmetry
breaking to occur, as can be seen in~\ref{fig:all_circuits}A. This circuit contains a positive autoregulation. As shown in~\cite{leon2016computational}, this allows for two stable states, making it possible for it to function as a memory device.

For NFBL (oscillator-type) circuits we observe 4 possible circuits: \emph{rob $\mapsto$ marA} a SCC by itself; \emph{soxS $\mapsto$ fur} in the \emph{soxS} SCC; and \emph{gadX $\mapsto$ hns} and \emph{cspA   $\mapsto$ hns} in the \emph{ph} SCC. All of these are autoregulated, but the autoregulations in \emph{gadX~$\mapsto$~hns} (Fig.~\ref{fig:all_circuits}) in fact makes it a Smolen oscillator, the more robust type oscillator studied in Ref.~\cite{stricker2008fast}.

There are also three pairs of nodes that can show various behaviors,
since they can send various types of regulation message between
them.  For example, \emph{crp} can send an activation or
repression signal to \emph{fis}, which means that \emph{crp-fis} can
be an MR circuit (toggle-switch type) or a NFBL circuit (an oscillating
type), possibly even a Smolen oscillator given its autoregulations. Similarly, for \emph{fnr-arcA} in the \emph{soxS}
SCC, which can be either. Lastly \emph{gadW-gadX} in the \emph{ph}
SCC, in which both genes send both activating and repressing signals,
can be a MR, a NFBL or a PAR feedback loop such as in a "lock-on"
circuit; on top of this, one of the possible NFBL configurations
includes a Smolen oscillator.

\begin{figure}
    \centering
    \includegraphics[width=.9\linewidth]{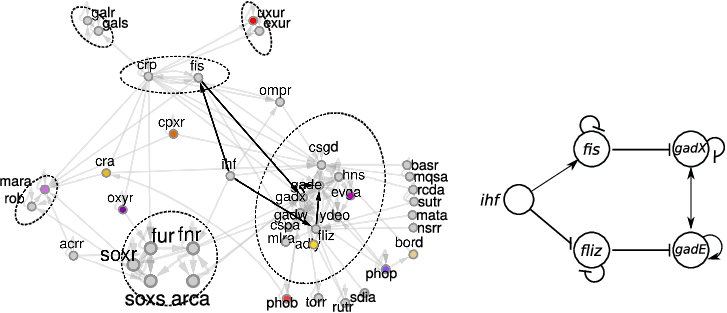}
    \caption{\textbf{FFF circuit in \emph{E. coli}.} FFF circuits connecting the main SCC \emph{crp-fis} to the \emph{pH}-SCC. On the right we show the isolated circuit.}
    \label{fig:fff circuits}
\end{figure}

More surprisingly, we found FFF circuits, shown in Fig.~\ref{fig:fff circuits}, that connect the \emph{pH} SCC to the master regulator \emph{crp-fis} SCC and, through two different paths, to the
\emph{soxS} SCC, remarkably, the three FFFs are regulated by the same
clock: \emph{ihfAB}. It should be noted that for the three FFF
circuits, the underlying feedback loop is a double positive
autoregulation (PAR) feedback loop, which works as a bistable
\emph{lock-on} circuit; however, the circuits actually inhibit PAR.

\SIsection{Testing the significance of network structures by comparison to randomized networks}
\label{sec:SI:statisticalAnalysis}

In order to be biologically meaningful, the network structures found with ComSym should be statistically significant compared to the results obtained from randomized graphs. To test this, we created an ensemble of 100 random networks, one corresponding to the characteristics of the \emph{E.~coli} network, and another for \emph{B. subtilis}.

\SIsubsection{Procedure}

Random networks were generated through a configuration model~\cite{molloy1995critical}, that is, a random null
model that preserves the in and out degrees of all network nodes,
along with the type of edges, while individual connections between
node pairs are randomized. In essence, we rewire the original network
to randomize its edges, without adding new edges or nodes, at the end,
the degree (both \emph{in} and \emph{out}) of each node remains
unchanged. This is done by repeatedly choosing two edges from the
original network at random and flipping their target nodes (i.e., edges
$A\mapsto B$ and $X\mapsto Y$ become $A\mapsto Y$ and $X\mapsto B$),
doing so repeatedly a number of times (10000 in our case). Self-edges
are not included in the flipping. We then perform the same analysis on
these random networks to compare the average structures across both
ensembles to what was found on the \emph{E.~coli} and \emph{B.~subtilis} networks, respectively. The significance of a given property (e.g~the number of appearances of a specific structure)
is measured by its \emph{Z} score from the mean values obtained from
each ensemble; see Table~\ref{tab:rand} for the full breakdown.


\SIsubsection{Results}

\begin{table*}[h]
 {\small
 \begin{tabular}{llll|lll}
     & \multicolumn{3}{c}{\cellcolor{myEcoliColor} \bf \emph{E.~coli}} & \multicolumn{3}{|c}{\cellcolor{myBsubColor} \bf \emph{B. subtilis}} \\
     & \bf\textit{N}$_{\rm real}$ & $N_{\rm rand} \pm {\rm SD}$ & $Z-score$ & $N_{\rm real}$ & $N_{\rm rand} \pm {\rm SD}$ & $Z-score$ \\
    \midrule
    & \multicolumn{6}{c}{\bf SCCs} \\
    \midrule
    Number of SCCs & 6 & 1.61 $\pm$ 0.74 & 5.93 & 4 & 1.2 $\pm$ 0.43 & 6.51 \\ 
    Average SCC size & 4 & 10.03 $\pm$ 7.94 & -0.76 & 4.75 & 25 $\pm$ 11.7 & -1.73 \\ 
    \midrule
    & \multicolumn{6}{c}{\bf Circuits} \\
    \midrule
    Total number of circuits & 13 & 2.02 $\pm$ 1.39 & 7.9 & 8 & 3.95 $\pm$ 1.98 & 2.05 \\ 
    Toggle-switch & 3 & 0.37 $\pm$ 0.58 & 4.53 & 2 & - & Inf \\ 
    Toggle-switch/Osc & 2 & 0.13 $\pm$ 0.34 & 5.5 & 0 & 0.6 $\pm$ 0.94 & -0.64 \\ 
    Oscillator & 5 & 0.7 $\pm$ 0.77 & 5.58 & 2 & 2.97 $\pm$ 1.86 & -0.52 \\ 
    Lock-on & 0 & 0.45 $\pm$ 0.64 & -0.7 & 4 & --- & Inf \\ 
    Lock-on/Osc & 0 & 0.17 $\pm$ 0.38 & -0.45 & --- & --- & ---- \\ 
    FFF & 3 & 0.2 $\pm$ 0.6 & 4.67 & 0 & 0.38 $\pm$ 0.69 & -0.55 \\ 
    \midrule
    & \multicolumn{6}{c}{\bf Cycles} \\
    \midrule
    Number of Cycles & 41 & 40.72 $\pm$ 70.9 & 0.00 & 48 & 2037.84 $\pm$ 5560 & -0.36 \\ 
    Average Cycle Length & 3.68 & 8.35 $\pm$ 3.53 & -1.32 & 4.88 & 15.49 $\pm$ 4.43 & -2.4 \\ 
    \hline\\
  \end{tabular}
  }
\small\sf\centering
 \caption{\textbf{Statistical significant structures in both GRNs}. 100 random networks with identical in- and out-degrees of all nodes and identical edge types were generated, for both of \emph{E.~coli}'s and \emph{B. subtilis}'s distributions. From these randomly generated networks we calculated the average and standard deviation quantities ($N_{rand} \pm SD$) to compare with the observed ones in the real distribution. \emph{Z}-scores of each quantity were calculated to show the statistical significance of the findings.}
  
  \label{tab:rand}
\end{table*}

Our analysis reveals rich structures in the core of both
bacteria's networks. Not only is the number of SCCs significant (with
a \emph{Z}-score > 5 for both bacterial networks), but also the
connections between the SCCs form an interesting structure for both
bacteria. We can see this structure for \emph{E.~coli} 
and for \emph{B. subtilis} 
in Fig.~\ref{fig:hist}B. In both networks, one
\emph{"central"} SCC regulates all the other SCCs in a
\emph{feed-forward} structure. We see two of these feed-forward structures in
\emph{E.~coli} and one in \emph{B. subtilis}, all involving the
\emph{central} SCC as the source. Most of the randomized networks, for
both types of bacteria, have only one or two SCCs in their core, that
is, basically no structure at all (see Fig.~\ref{fig:hist}). Only 12
of the randomized \emph{E.~coli} networks show more than 2 SCCs, and
only one of them shows 4 SCCs, while the ensemble of randomized
\emph{B.~subtilis} networks contains only one network with 3 SCCs.

The average SCC size is not highly significant (according to its
\emph{Z}-score alone). We observe that the \emph{size} of the SCCs
decreases as the \emph{number} of the components increases, which means that the more components found in these random networks, the smaller they were.
For both \emph{E.~coli}'s and \emph{B. subtilis}' GRNs, however, we found as many as 6 and 4 SCCs, some with sizes of 11 and 13
nodes, respectively. This is in sharp contrast to any structure observed in the randomized networks. For
\emph{E.~coli}, the average size of the SCC, if the \emph{only one} component is
present, is 16.76 however, if there is \emph{more than one} SCC, the
average size now becomes 6.79; the biggest component has a size of 31
nodes for random networks with it being the only one for that particular network, on the other hand for random networks with three components its only 18 nodes in
size. The same trend is true for \emph{B. subtilis}.

Aside from this large-scale structure, we also found a variety of
logic circuits on the smaller scale, with 12 circuits in total
(\emph{Z}-score = 7.9) for \emph{E.~coli} and 8 circuits
(\emph{Z}-score = 2.05) for \emph{B. subtilis}. The appearance of
circuits is very specific. In the case of \emph{E.~coli}, the
randomized networks show no preference towards any particular type of
circuit: the expected count numbers for all types of circuits are
lower than 1. In the real network the count numbers of the observed
circuit types were much higher, with significant \emph{Z}-scores, see
Table~\ref{tab:rand}. In the \emph{B. subtilis} random networks, there
is a bias towards the presence of oscillators of which we found 2 in
the real network. We also found 2 toggle-switches and 4 lock-ons,
although none of the two were found in the randomized networks.

\SIsection{Algorithms and pseudocode}
\SIsubsection{Minimal Balanced Coloring algorithm}
\label{SI:Algorithm}

This process of partitioning a network into sets of fibers with
isomorphic input tress is {equivalent to finding the minimal
balanced coloring or balanced equivalence relations of the network~\cite{aldis2008polynomial, kamei2013computation}. A balanced coloring of a network occurs when the colors assigned on the nodes are such that they are balanced, which is to say that nodes in a given color receive the same number of inputs from each color they receive. 
This corresponds to an equivalence relation on the set of nodes~\cite{cardon1982partitioning}, partitioning the network into balanced equivalence relations, hence its name~\cite{aldis2008polynomial, kamei2013computation}.}
It is worth mentioning that undirected networks can have multiple balanced colorings including some exotic ones~\cite{golubitsky2004some, golubitsky2016rigid}, however, for the case of directed networks, given that the order of signals is more clearly defined, this does not seem to be an issue.

Our algorithm is based on a minimal balanced coloring
algorithm proposed by Aldis and Kamei~\cite{aldis2008polynomial, kamei2013computation, cardon1982partitioning, morone2020fibration}.
The algorithm starts by giving all the nodes the same color, except
for the nodes with no input, since biologically there is no reason for
them to be synchronous. Then the color of the nodes are recalculated:
nodes are given the same color if they receive the same amount of
inputs for every color they receive. For example, two genes that both
receive one blue and one red input (just to say some colors as
examples) are given a same color. If a third node receives two blue inputs and one red input, it is given a different color. By changing the
colors, the color inputs of the nodes change as well. The colors are
then re-calculated again based on the new color-input relations. Once
the coloring is stable, the program is stopped. At this point, we have
found the coloring partition where each color is a fiber. The code
outputs the minimal balanced coloring, which represents the maximal
fibration symmetry of the graph. This means that there are other
coloring partitions that are not minimal balanced, for instance, we can always break a single fiber
into two different colors, and decrease the symmetry, or equivalently,
increase the number of colors, and this new coloring will not be the
minimal. See~\cite{morone2020fibration} and its Supplementary
Information for more information, as well as~\cite{monteiroAlgorithm}. Our implementation of the Minimal Balanced Coloring algorithm can be found at \url{https://github.com/makselab/MinimalTRNCodes}.  

{One inadvertent advantage of these biological networks that show fibration symmetries and not automorphisms is the computational cost of calculating them. 
The {\it graph isomorphism problem} attempts to determine whether two finite graphs are isomorphic. It is a very active field of research as it is an immensely complicated problem to solve generally, Babai~\cite{babai2016graph} has shown that graph isomorphisms can be solved in quasi-polynomial time; however, no polynomial-time algorithms are known today. It is still not clear if Graph Isomorphisms belongs to NP-complete~\cite{babai2016graph}. Fortunately for us, however, since graph automorphisms are too restrictive for these real biological networks we only need to look on input tree isomorphisms instead, which are much less computationally costly: even for a pair of infinite input trees, it suffices to show the isomorphism up to $N_G - 1$ layers of the trees to determine the isomorphism~\cite{norris1995universal}. Furthermore, the balanced coloring algorithm scales only with polynomial time~\cite{aldis2008polynomial, kamei2013computation, cardon1982partitioning}.}

\SIsubsection{ComSym Method}

The entire code can be found at \url{https://github.com/luisalvarez96/MinimalTRN}.

\paragraph{Steps I to III}

Although the column {\bf N}[{\it "FiberSize"}] is not directly used in the below pseudocode Algorithm~\ref{pseudo1}, it is a necessary output for the code in order to separate the nodes that belong to a fiber from those that do not. This is because after {\bf Step I}, having collapsed all fibers into a single node, there are no repeated values in {\bf N}$_{col}$[{\it "FiberId"}], and hence without having the column {\bf N}$_{col}$[{\it "FiberSize"}] it would be impossible to tell which nodes belong to a fiber. This is because the {\it Minimal Balanced Coloring} algorithm returns a {\it "FiberId"} value {\it for each} node, even for nodes with no fiber (or single-fiber nodes).

With the output of this code, we can group nodes according to {\bf N}$_{min}$[{\it "SCCId"}], to determine the nodes that belong to a SCC bigger than one, determine how many SCCs are there, and study the interaction between them to finally reveal the {\it large-scale structure} of the minimal network.

\begin{algorithm}[H]
\caption{ComSym Method: Steps I-III.}
\label{pseudo1}
{\bf Notation}: 
\begin{itemize}
    \item {\bf N}[{\it "Label", "FiberId"}] referes to a list, or data frame, {\bf N} with columns {\it "Label"}, and {\it "FiberId"}.
    \item {\bf N}[{\it "Label"}] refers to column {\it "Label"} of data frame {\bf N}.
\end{itemize}

\textbf{Input:} A list of edges {\bf E}[{\it "Source", "Target", "Type"(optional)}], from graph $G = (N, E)$, where columns stand for: source node label, target node label, and type of interaction (activation, inhibition, etc.).

\textbf{Output:} A list {\bf N$_{min}$} of nodes and a list {\bf E$_{min}$} of edges (forming the minimal graph $G_{min} = (N_{min}, E_{min})$ of graph $G$).
\newline
\begin{algorithmic}[1] 
\Statex {\bf Obtaining the Fibers (colors):}
\State {\bf N}[{\it "Label", "FiberId"}] $\leftarrow$ {\it Minimal Balanced Coloring}({\bf E}) (see Section~\ref{SI:Algorithm} and SI from~\cite{morone2020fibration}).
\Statex Each row in {\bf N} corresponds to a gene, with columns {\it Label}: gene name for each node; and {\it FiberId}: the assigned fiber (or color) for the corresponding node.
    
\State Add column {\it FiberSize} (size of the corresponding node's fiber) to {\bf N}: {\bf N}[{\it "Label", "FiberId", "FiberSize"}].

\Statex   

\Statex {\bf Step I: Collapsing (symmetry fibration)}
\State Initialize an empty data frame {\bf N$_{col}$} (for collapsed nodes).
\For{each fiber in {\bf N}[{\it "FiberId"}]}
    \State Select one node (row) in {\bf N} from that fiber and append its entire row to {\bf N$_{col}$}. 
\EndFor

\State {\bf E$_{col}$} $\leftarrow$ From {\bf E} select rows {\bf if} {\bf E}[{\it "Target"}] in {\bf N$_{col}$}[{\it "Label"}] (for collapsed edges).
\For{each edge in {\bf E$_{col}$}}
    \State \textbf{if} {\bf E$_{col}$}[{\it "Source"}] not in {\bf N$_{col}$}[{\it "Label"}] {\bf then}:
        \State  \qquad Find the row in {\bf N$_{col}$} where {\bf N$_{col}$}[{\it "FiberId"}] == {\it FiberId} from {\bf E$_{col}$}[{\it "Source"}]. 
        \State \qquad Replace {\bf E$_{col}$}[{\it "Source"}] with {\bf N$_{col}$}[{\it "Label"}] from the found row. 
\EndFor

\Statex

\Statex {\bf Step II: Pruning ($k_{out}$-core decomposition)}
\State Add column {\it Outdegree} (outdegree for the corresponding gene, self-loops not included) to {\bf N}$_{col}$: {\bf N}$_{col}$[{\it "Label", "FiberId", "FiberSize", "Outdegree"}]
\State {\bf N}$_{min} \leftarrow$ From {\bf N}$_{col}$ select rows {\bf if} {\bf N$_{col}$}[{\it "Outdegree"}] > 0
\State {\bf E}$_{min} \leftarrow$ From {\bf E}$_{col}$ select rows {\bf if} {\bf E$_{col}$}[{\it "Target"}] in {\bf N$_{min}$}[{\it "Label"}]
\State {\bf while} number of rows of {\bf N}$_{min}$ is not equal to number of rows of {\bf N}$_{col}$ {\bf do}:
    \State \qquad Recalculate {\bf N$_{min}$}[{\it "Outdegree"}]
    \State \qquad {\bf N}$_{col} \leftarrow$ {\bf N}$_{min}$
    \State \qquad {\bf E}$_{col} \leftarrow$ {\bf E}$_{min}$
    \State \qquad {\bf N}$_{min} \leftarrow$ From {\bf N}$_{min}$ select rows {\bf if} {\bf N$_{min}$}[{\it "Outdegree"}] > 0
    \State \qquad {\bf E}$_{min} \leftarrow$ From {\bf E}$_{min}$ select rows {\bf if} {\bf E$_{min}$}[{\it "Target"}] in {\bf N$_{min}$}[{\it "Label"}]

\Statex

\Statex {\bf Step III. Large-scale structure of the minimal network (SCCs)}
\State Add column {\it SCCId} (numerical identifier for the SCC the corresponding node belongs to) to {\bf N}$_{min}$: {\bf N}$_{min}$[{\it "Label", "FiberId", "FiberSize", "Outdegree", "SCCId"}]

\Statex

\State {\bf return} {\bf N}$_{min}$[{\it "Label", "FiberId", "FiberSize", "Outdegree", "SCCId"}] 
\\and {\bf E}$_{min}$[{\it "Source", "Target", "Type"}]

\end{algorithmic}
\end{algorithm}

\paragraph{Step IV}
To find the logic circuits, we run a modified version of the algorithm developed
by Leifer {\it et al.} in Ref.~\cite{leifer2020circuits} looking for the induced subgraphs of the network whose connectivity is identical to the logic circuits we are looking for. The algorithm consists of essentially two steps. 

\begin{algorithm}[H]
\caption{ComSym Method: Steps IV.}
\label{pseudo2}
{\bf Notation:} same as in previous Algorithm~\ref{pseudo1}\\
\textbf{Input:} A list of edges {\bf E}$_{min}$[{\it "Source", "Target", "Type"(optional)}].

{\bf Initialize:} An adjacency matrix {\bf Adj} for each circuit to look for.

\textbf{Output:} A list {\bf C} of circuits.
\newline
\begin{algorithmic}[1] 
\State {\bf E} $\leftarrow$ From {\bf E} select rows {\bf if} {\bf E}[{\it "Source"}] is not equal to {\bf E}[{\it "Target"}]
\State {\bf E} $\leftarrow$ From {\bf E} remove duplicated rows
\State {\bf C} $\leftarrow$ search for subgraphs of {\bf E}$_{min}$ isomorphic to {\bf Adj}
\State {\bf C} $\leftarrow$ From {\bf C} select only induced subgraphs
\State {\bf return C} 

\end{algorithmic}
\end{algorithm}

The {\bf Adj} matrices for the circuits correspond to:
\begin{equation*}
AR = \begin{bmatrix}
    0 & 1 \\
    1 & 0 
\end{bmatrix}
\quad
FFF = \begin{bmatrix}
    0 & 0 & 0 & 0 & 0\\
    1 & 0 & 0 & 0 & 1\\
    1 & 0 & 0 & 1 & 0\\
    0 & 1 & 0 & 0 & 1\\
    0 & 0 & 1 & 1 & 0\\
\end{bmatrix}
\end{equation*}

For circuits that originate from the duplication of an auto-regulated (AR) gene or a FeedForward fiber (FFF) building block, respectively. AR circuits can then be further classified into a toggle-switch, a lock-on circuit, or an oscillator depending on their type of edges. A toggle-switch corresponds to both edges being repressive, an oscillator to one negative and one positive and a lock-on for both edges as activations. In actuality, for both studied GRNs, the first two lines in Algorithm~\ref{pseudo2} are required since virtually no circuits are exactly identical to the "neat" and nicely behaving circuits design and implemented synthetically. Perhaps it should come as no surprise that actual in-vivo reality is more complicated. 
Some of the observed circuits, as a result, have a slightly different topology due to self-regulations of some genes in the circuits or due to multiplicity of "paralel" edges, so their specific dynamical behavior requires further study. 

\paragraph{Step V}
Both networks studied
here are small enough to be analyzed using an algorithm
loosely based on \emph{Johnson's
  algorithm}~\cite{johnson1977efficient} and shown on Algorithm \ref{pseudo3}. It starts by breaking the
network into SCCs and searches for cycles within each one. Each
component is analyzed by 1) enumerating the nodes in the SCC, 2)
choosing one node (the initial/final node of the cycle), 3) looking
for the outgoing neighbors of this node, and 4) looking for all the
simple paths (paths without repeating nodes) back from each neighbor
to the initial/final node.

\begin{algorithm}[H]
\caption{ComSym Method: Steps V.}
\label{pseudo3}
{\bf Notation:} same as in previous Algorithm~\ref{pseudo1}\\
\textbf{Input:} {\bf N}$_{min}$[{\it "Label", "FiberId", "FiberSize", "Outdegree", "SCCId"}] \\   and {\bf E}$_{min}$[{\it "Source", "Target", "Type"(optional)}].

\textbf{Output:} A list {\bf C} of cycles.
\newline
\begin{algorithmic}[1] 
\State Identify all {\it SCCId}'s for the SCCs from {\bf N}$_{min}$[{\it "SCCId"}]
\State Initialize empty list {\it Cycles}
\For{every {\it SCCId} in SCCs}
    \State enumerate all nodes in {\it SCCId} (order is arbitrary, results will be the same regardless )
    \State {\it NodeCount} $\leftarrow$ Number of nodes in list of enumerated nodes
    \While{{\it NodeCount} is not equal to 1}
        \State {\it node} $\leftarrow$ first node from list of enumerated nodes
        \State {\it neighbors} $\leftarrow$ out-neighbors from node
        \State Initialize empty list {\it paths}
        \For{{\it neighbor} in {\it neighbors}}
            \State find all simple paths from {\it neighbor} to {\it node} in {\bf E}$_{min}$
            \State append paths from {\it neighbor} to {\it node} to {\it paths} 
        \EndFor
        \State Add columns {\it SCC} ({\it SCCId} of current SCC) and {\it Lenght} (length of each path) to {\it paths} 
        \State Append {\it paths} to  {\it Cycles}  
        \State Remove {\it node} from list of enumerated nodes
        \State {\it NodeCount} $\leftarrow$ Number of nodes in list of enumerated nodes
    \EndWhile
\EndFor
\State {\bf C} $\leftarrow$ From {\bf C} remove duplicated rows
\State {\bf return C} 

\end{algorithmic}
\end{algorithm}

\bibliography{main}

\end{document}